\documentclass[11pt,a4paper]{article}
\pdfoutput=1

\usepackage{jheppub}
\usepackage{amsmath}
\usepackage{amssymb}
\usepackage{graphics}
\usepackage[active]{srcltx}
\usepackage{pdfsync}
\usepackage{shuffle}
\usepackage{slashed}
\usepackage{hyperref}
\usepackage{subfigure}

\newcommand{\bea}{\begin{eqnarray}}
\newcommand{\eea}{\end{eqnarray}}
\newcommand{\nn}{\nonumber \\}

\def\W #1{\widetilde{#1}}

\def\eref#1{(\ref{#1})}

\allowdisplaybreaks

\title{On soft factors and transmutation operators}
\author[a]{Fang-Stars Wei} \author[a]{Kang Zhou}

\affiliation[a]{Center for Gravitation and Cosmology, College of Physical Science and Technology, Yangzhou University,\\
No.180, Siwangting Road, Yangzhou, 225009, P.R. China.}

\emailAdd{mx120220339@stu.yzu.edu.cn} \emailAdd{zhoukang@yzu.edu.cn}

\date{\today}
\abstract{The well known soft theorems state the specific factorizations of tree level gravitational (GR) amplitudes at leading, sub-leading
and sub-sub-leading orders, with universal soft factors. For Yang-Mills (YM) amplitudes, similar factorizations and universal
soft factors are found at leading and sub-leading orders. Then it is natural to ask if the similar factorizations and soft factors exist at higher orders. In this note, by using transformation operators proposed by Cheung, Shen and Wen, we reconstruct the known soft factors of YM and GR amplitudes, and prove the nonexistence of higher order soft factor of YM or GR amplitude which satisfies the universality.}


\begin{document}

\maketitle \flushbottom

\section{Introduction}
\label{sec-intro}

In recent years, the investigation on soft theorems of scattering amplitudes has been an active area of research, leading to remarkably insights and applications ranging from gauge theory and gravity, to various effective field theories (EFTs).
Soft theorems describe the universal behaviors of amplitudes when one or more external momenta are taken to near zero.
Historically, they were originally discovered for photons and gravitons, at tree level \cite{Low:1958sn,Weinberg:1965nx}. In 2014, the soft behaviors of tree gravitational (GR) and Yang-Mills (YM) amplitudes were extended to higher-orders \cite{Cachazo:2014fwa,Casali:2014xpa,Schwab:2014xua,Afkhami-Jeddi:2014fia,Zlotnikov:2014sva}, by using modern technics beyond Feynman diagrams, like Britto-Cachazo-Feng-Witten (BCFW) \cite{Britto:2004ap,Britto:2005fq} recursion relation and Cachazo-He-Yuan (CHY) formula \cite{Cachazo:2013gna,Cachazo:2013hca, Cachazo:2013iea, Cachazo:2014nsa,Cachazo:2014xea}. Subsequently, soft theorems were further studied in a wider range including string theory and the loop level \cite{Bern:2014oka,He:2014bga,Cachazo:2014dia,Bianchi:2014gla,Sen:2017nim}. Meanwhile, it turns out that tree amplitudes can be constructed by solely exploiting soft behaviors, with out the aid of a Lagrangian or Feynman rules, see in progresses in \cite{Nguyen:2009jk,Boucher-Veronneau:2011rwd,Rodina:2018pcb,Ma:2022qja,Cheung:2014dqa,Cheung:2015ota,Luo:2015tat,Elvang:2018dco,
Zhou:2022orv,Wei:2023yfy,Hu:2023lso,Du:2024dwm}.

The soft limit can be achieved by rescaling the external momentum $k_i$ carried by particle $i$ as $k_i\to\tau k_i$, then take the limit $\tau\to0$. For gravity and gauge theory, soft theorems state that in the soft limit the full $n$-point amplitude factorizes into a soft factor,
as well as a $(n-1)$-point sub-amplitude. For instance, the $n$-point GR amplitude factorizes as
\bea
{\cal A}_{n}\,\to\,\Big(\tau^{-1}\,S^{(0)_i}_h+\tau^0\,S^{(1)_i}_h+\tau\,S^{(2)_i}_h\Big)\,{\cal A}_{n-1}+{\cal O}(\tau^2)\,,~~~~\label{softtheo}
\eea
where ${\cal A}_{n-1}$ is the sub-amplitude of ${\cal A}_n$, generated from ${\cal A}_n$ by removing the soft external graviton.
The operators $S^{(0)_i}_h$, $S^{(1)_i}_h$, $S^{(2)_i}_h$ are called soft factors, at leading, sub-leading, and sub-sub-leading orders respectively,
their precise forms can be seen in \cite{Cachazo:2014fwa,Schwab:2014xua,Afkhami-Jeddi:2014fia,Zlotnikov:2014sva}.
These factors are universal, namely, their forms are valid for arbitrary $n\geq4$.
For GR amplitudes, $S^{(0)_i}_h$, $S^{(1)_i}_h$, $S^{(2)_i}_h$ in \eref{softtheo} are all already known soft factors. For YM amplitudes, we have
two known soft factors $S^{(0)_i}_g$ and $S^{(1)_i}_g$, at leading and sub-leading orders. Since each amplitude can always be expanded
to a series with respect to $\tau$,
\bea
{\cal A}_n(\tau)=\sum_{a=0}^{+\infty}\,\tau^{a-1}\,{\cal A}_n^{(a)_i}\,,~~\label{expan-general}
\eea
it is natural to ask, can higher order terms ${\cal A}_n^{(a)_i}$ factorize as ${\cal A}^{(a)_i}_n=S^{(a)_i}\,{\cal A}_{n-1}$ in the soft limit?

Formally, the answer is extremely trivial, since one can define $S^{(a)_i}\equiv{\cal A}_n^{(a)_i}/{\cal A}_{n-1}$ in \eref{expan-general},
then the factorization behavior ${\cal A}^{(a)_i}_n=S^{(a)_i}\,{\cal A}_{n-1}$ holds at any order. However, such formal factorization
does not lead to any physical insight. Thus, it is necessary to impose further physical criteria as the constraint on soft factors. The most natural candidate of such criteria is the universality, which is satisfied by known soft factors for GR and YM amplitudes, as well as the distinct soft behavior called Adler zero for various EFTs. Thus, we are interested in the existence of universal soft factors $S^{(a)_i}_h$ and $S^{(a)_i}_g$
at higher orders, with $a\geq3$ for GR and $a\geq2$ for YM, respectively.

In this note, with the help of transmutation operators which connect amplitudes of different theories together \cite{Cheung:2017ems,Zhou:2018wvn,
Bollmann:2018edb}, we prove that there is no universal soft factor can be found at higher order. In other words, for GR amplitudes, the only soft factors satisfy the universality are those $S^{(a)_i}_h$ with $a=0,1,2$. Meanwhile, soft factors for YM amplitudes are $S^{(a)_i}_g$ with $a=0,1$. Our method is as follows. The GR, YM, bi-adjoint scalar (BAS) amplitudes are linked by the transmutation operator ${\cal T}[1,\cdots,n]$. On the other hand,
as will be explained in section \ref{subsec-BAS}, it is straightforward to figure out the leading soft factor
of BAS amplitudes, and observe that no soft factor compatible with universality can be found at higher order. Based on transmutation relations and the known leading soft behavior of BAS amplitudes, we can establish equations which allow us to solve potential $S^{(a)_i}_g$ and $S^{(a)_i}_h$. We then find all solutions of $S^{(a)_i}_g$ and $S^{(a)_i}_h$, coincide with those in literatures \cite{Cachazo:2014fwa,Casali:2014xpa,Schwab:2014xua,Afkhami-Jeddi:2014fia,Zlotnikov:2014sva}, and prove the nonexistence of solution at higher order. As a byproduct, we also
clarify that the consistent soft factors $S^{(a)_i}_h$ with $a=1,2$ found in literatures \cite{Cachazo:2014fwa,Schwab:2014xua,Afkhami-Jeddi:2014fia,Zlotnikov:2014sva} only hold for amplitudes of standard Einstein gravity. For the extended theory that Einstein gravity couples to $2$-form and dilaton field, they are spoiled.

The note is organized as follows. In section \ref{sec-preparations}, we give a brief review for soft behavior of tree BAS amplitudes,
as well as transmutation operators. Then, in section \ref{sec-YM} we rederive soft factors of YM amplitudes at leading and sub-leading orders,
and prove the nonexistence of higher order soft factors. Subsequently, in section \ref{sec-GR}, we rederive soft factors of GR amplitudes at
leading, sub-leading, sub-sub-leading orders, and prove the nonexistence of higher order ones. Finally, a brief summary will be presented in
section \ref{sec-summary}.

\section{Back ground}
\label{sec-preparations}

In this section, we give a rapid review for necessary background, including the soft behavior of BAS amplitudes, as well as transmutation operators
proposed in \cite{Cheung:2017ems}.

\subsection{Soft behavior of BAS amplitudes}
\label{subsec-BAS}

\begin{figure}
  \centering
  \includegraphics[width=6cm]{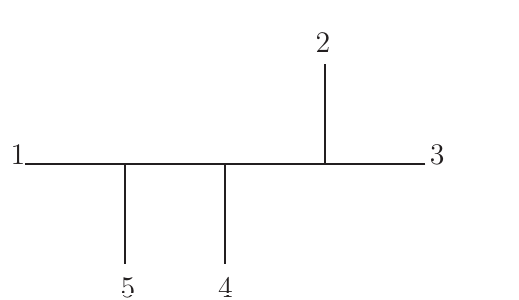}
   \includegraphics[width=6cm]{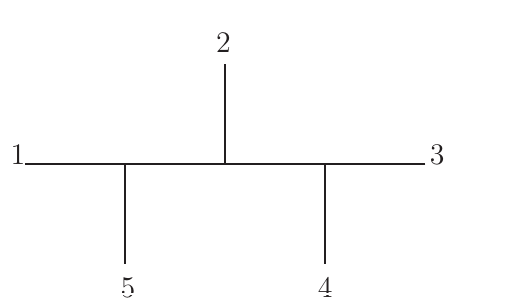}  \\
  \caption{Two $5$-point diagrams}\label{5p}
\end{figure}

The bi-adjoint scalar (BAS) amplitudes describe scattering of massless scalars, with cubic interactions. In this paper, we are interested in the double ordered partial BAS amplitudes at the tree level. Each $n$-point double ordered amplitude ${\cal A}_{\rm BAS}(\vec{\pmb\sigma}|\vec{\pmb\sigma}'_n)$ carries two orderings encoded as $\vec{\pmb\sigma}$ and $\vec{\pmb\sigma}'_n$, and is simultaneously planar with respect to both two orderings.
Here we give a $5$-point example. The first diagram in Figure.\ref{5p} contributes to the amplitude ${\cal A}_{\rm BAS}(1,2,3,4,5|1,4,2,3,5)$, since it is compatible with both two orderings $1,2,3,4,5$ and $1,4,2,3,5$. Meanwhile, the second diagram violates the ordering
$1,4,2,3,5$, thus is forbidden. One can verify that the first diagram in Figure.\ref{5p} is the only candidate satisfies two orderings simultaneously, thus the amplitude ${\cal A}_{\rm BAS}(1,2,3,4,5|1,4,2,3,5)$ reads
\bea
{\cal A}_{\rm BAS}(1,2,3,4,5|1,4,2,3,5)={1\over s_{23}}{1\over s_{51}}\,,
\eea
up to an overall sign. Here the Mandelstam variable $s_{i\cdots j}$ is defined as
\bea
s_{i\cdots j}\equiv k_{i\cdots j}^2\,,~~~{\rm with}~k_{i\cdots j}\equiv\sum_{a=i}^j\,k_a\,,~~~~\label{mandelstam}
\eea
where $k_a$ is the momentum carried by the external leg $a$.

In double ordered BAS amplitudes, each interaction vertex features antisymmetry for lines attached to it. In other words, swamping two lines $a$ and $b$ at a vertex creates a $-$ sign, therefore BAS amplitudes with different orderings carry different overall $\pm$ sign. To simplify the description of soft behavior, we choose the overall sign to be $+$ if two orderings carried by the BAS amplitude are the same. For example, the amplitude ${\cal A}_{\rm BAS}(1,2,3,4|1,2,3,4)$ carries the overall sign $+$ under our convention. The sign for amplitudes with two different orderings can be generated from the above reference one by counting flippings, via the diagrammatic technic proposed in \cite{Cachazo:2013iea}.

Due to the definition of tree BAS amplitudes introduced above, it is direct to observe the leading soft behavior of BAS amplitude ${\cal A}_{\rm BAS}(1,\cdots,n|\vec{\pmb\sigma}_n)$. Take the external scalar $i$ to be the soft particle, with $k_i^{\mu}\sim \tau k_i^{\mu}$ and $\tau\to 0$, then the propagators $1/s_{i(i-1)}$ and $1/s_{i(i+1)}$ become divergent in such soft limit, therefore
\bea
{\cal A}^{(0)_i}_{\rm BAS}(1,\cdots,n|\vec{\pmb\sigma}_n)&=&{1\over\tau}\,S^{(0)_i}_s\,{\cal A}_{\rm BAS}(1,\cdots,i-1,i+1,\cdots,n|\vec{\pmb\sigma}_n\setminus i)\,.~~~\label{for-soft-fac-s}
\eea
Throughout this paper, we use the superscript $(a)_i$ to denote the contribution at the $a^{\rm th}$ order, with the particle $i$ taken to be soft.
In the above, the leading soft factor $S^{(0)_i}_s$ is given by
\bea
S^{(0)_i}_s=\left(\,{\delta_{i(i+1)}\over s_{i(i+1)}}+{\delta_{(i-1)i}\over s_{(i-1)i}}\,\right)\,,~~~~\label{soft-fac-s-0}
\eea
where the symbol $\delta_{ij}$ is determined by positions of $i$ and $j$ in the ordering $\vec{\pmb\sigma}_n$. If $i$, $j$ are not adjacent to each other, $\delta_{ij}=0$. If $i$ and $j$ are two adjacent elements, we have $\delta_{ij}=1$ for $i\prec j$, $\delta_{ij}=-1$ for $j\prec i$. The notation $a\prec b$ for a given permutation including elements $a$, $b$ means the position of $a$ is on the left side of the position of $b$.

The higher order soft behaviors can also be analysed by considering contributions from allowed Feynman diagrams. However, suppose we formally express them as
\bea
{\cal A}^{(a)_i}_{\rm BAS}(1,\cdots,n|\vec{\pmb\sigma}_n)&=&\tau^{(a-1)}\,S^{(a)_i}_s\,{\cal A}_{\rm BAS}(1,\cdots,i-1,i+1,\cdots,n|\vec{\pmb\sigma}_n\setminus i)\,,
\eea
the soft factors $S^{(a)_i}_s$ with $a\geq1$ do not satisfy the universality. For example, the $4$-point BAS amplitudes only have the leading order term, while the higher-point amplitudes receives contribution from higher orders. In this sense, for tree BAS amplitudes under consideration, the expected universal soft factor only exist at the leading order.

\subsection{Transmutation operators}
\label{subsec-operator}

The transmutation operators proposed by Cheung, Shen and Wen connects tree amplitudes of different theories together, by transmuting amplitudes of one theory to those of another \cite{Cheung:2017ems}. In this paper, we only use the combinatorial operator ${\cal T}[1,\cdots,n]$, which turns GR amplitudes to ordered YM amplitudes, and also transmutes YM amplitudes to double ordered BAS amplitudes, namely,
\bea
{\cal A}_{\rm YM}(1,\cdots,n)&=&\W{\cal T}[1,\cdots,n]\,{\cal A}_{\rm GR}(\pmb{h}_n)\,,\nn
{\cal A}_{\rm BAS}(1,\cdots,n|\vec{\pmb\sigma}_n)&=&{\cal T}[1,\cdots,n]\,{\cal A}_{\rm YM}(\vec{\pmb\sigma}_n)\,.~~\label{trans-rela}
\eea
In the above, ${\pmb h}_n$ stands for the unordered set of $n$ external gravitons, and $\vec{\pmb\sigma}_n$ labels the ordered set of $n$ external gluons.
The polarization tensor of each graviton is decomposed as $\varepsilon_p^{\mu\nu}=\epsilon_p^\mu\W\epsilon_p^\nu$, where $\epsilon_p$
and $\W\epsilon_p$ with $p\in\{1,\cdots,n\}$ are two sectors of polarization vectors. The operator $\W{\cal T}[1,\cdots,n]$ is defined for polarization vectors $\W\epsilon_p$ (as will be explicitly explained soon), transmutes the unordered GR amplitude ${\cal A}_{\rm GR}(\pmb{h}_n)$ to the YM amplitude ${\cal A}_{\rm YM}(1,\cdots,n)$ with ordering $1,\cdots,n$, and the external gluons carry polarization vectors $\epsilon_p$.
The operator ${\cal T}[1,\cdots,n]$ is defined for polarization vectors $\epsilon_p$, transmutes the ordered YM amplitude ${\cal A}_{\rm YM}(\vec{\pmb\sigma}_n)$ to the double ordered BAS amplitude ${\cal A}_{\rm BAS}(1,\cdots,n|\vec{\pmb\sigma}_n)$ for scalars without any polarization. Notice that the relation in \eref{trans-rela} which turns GR amplitude to YM one only makes sense for the extended gravity theory that Einstein gravity is coupled to a $2$-form and dilaton field.

One of the explicit forms for the combinatorial operator ${\cal T}[1,\cdots,n]$ is given by
\bea
{\cal T}[1,\cdots,n]=\Big(\prod_{a=2}^{n-1}\,{\cal I}_{(a-1)an}\Big)\,\partial_{\epsilon_1\cdot\epsilon_n}\,,~~~~{\rm with}~\partial_x\equiv{\partial\over\partial x}\,,~~\label{definT-1}
\eea
where the insertion operator ${\cal I}_{bac}$ is defined as
\bea
{\cal I}_{bac}\equiv\partial_{\epsilon_a\cdot k_b}-\partial_{\epsilon_a\cdot k_c}\,.
\eea
The above ${\cal T}[1,\cdots,n]$ is defined for polarization vectors $\epsilon_p$. The dual operator $\W{\cal T}[1,\cdots,n]$, defined for polarization vectors $\W\epsilon_p$, can be obtained from ${\cal T}[1,\cdots,n]$ by replacing all $\epsilon_p$ with $\W\epsilon_p$.
As proved in \cite{Zhou:2018wvn,
Bollmann:2018edb}, the effect of insertion operator ${\cal I}_{bac}$ is to reduce the spin of particle $a$ by $1$, and insert it between $b$ and $c$ in the ordering. According to the above interpretation for insertion operator, the effect of the operator ${\cal T}[1,\cdots,n]$ in \eref{definT-1} can be understood as,
\begin{itemize}
 \item Generating two endpoints $1$ and $n$ of the ordering.
 \item Inserting the leg $2$ between $1$ and $n$.
 \item Inserting the leg $3$ between $2$ and $n$.
 \item Repeating the above procedure to insert other legs in turn, until the full ordering $1,\cdots,n$ is completed.
\end{itemize}
The above steps exhibit the process for generating the full ordering. Notice that we assume the order of performing differentials in \eref{definT-1} to be from right to left. However, since all insertion operators in \eref{definT-1} are algebraically commutative, we can rearrange the order as ${\cal I}_{(n-2)(n-1)n}{\cal I}_{(n-3)(n-2)n}\cdots{\cal I}_{23n}{\cal I}_{12n}$, and interpret the effects of them in the above way.

To create any desired ordering $\vec{\pmb\sigma}_n$, the corresponding formula of combinatorial operator ${\cal T}[\vec{\pmb\sigma}_n]$ is not unique, due to the interpretation for the insertion operator ${\cal I}_{bac}$. For instance, to generate the ordering $1,2,3,4$ one can chose
\bea
{\cal T}[1,2,3,4]={\cal I}_{234}\,{\cal I}_{124}\,\partial_{\epsilon_1\cdot\epsilon_4}\,,
\eea
which realizes the goal as
\begin{itemize}
 \item Generating two endpoints $1$ and $4$.
 \item Inserting the leg $2$ between $1$ and $4$.
 \item Inserting the leg $3$ between $2$ and $4$.
\end{itemize}
However, the different choice
\bea
{\cal T}[1,2,3,4]={\cal I}_{123}\,{\cal I}_{134}\,\partial_{\epsilon_1\cdot\epsilon_4}
\eea
is also correct, this operator generates the ordering $1,2,3,4$ as
\begin{itemize}
 \item Generating two endpoints $1$ and $4$.
 \item Inserting the leg $3$ between $1$ and $4$.
 \item Inserting the leg $2$ between $1$ and $3$.
\end{itemize}
Such freedom for choosing the combinatorial operator ${\cal T}[1,\cdots,n]$ will play the crucial role in subsequent sections.

The analogous operator ${\cal T}[\vec{\pmb\sigma}_m]$, where $\vec{\pmb\sigma}_m$ is the ordering among $m$ elements in the complete set $\{1,\cdots,n\}$, with $m<n$, also leads to the meaningful interpretation. For instance, one can define ${\cal T}[2,3,4]={\cal I}_{234}\partial_{\epsilon_2\cdot \epsilon_4}$. When acting on the $4$-point YM amplitude ${\cal A}_{\rm YM}(\vec{\pmb\sigma}_4)$ with external gluons encoded as $\{1,2,3,4\}$, the above operator ${\cal T}[2,3,4]$ turns gluons in the set $\{2,3,4\}$ to scalars, and generates the ordering $2,3,4$. The resulted amplitude ${\cal A}_{\rm YMS}(2,3,4;1|\vec{\pmb\sigma}_4)$ is known as
the Yang-Mills-scalar (YMS) amplitude that the gluon $1$ interacts with BAS scalars in $\{2,3,4\}$.
Similarly, the operators $\W{\cal T}[\vec{\pmb\sigma}_m]$ with $m<n$ transmutes GR amplitudes to Einstein-Yang-Mills (EYM) amplitudes those gravitons couple to gluons.

\section{Soft behavior of YM amplitudes}
\label{sec-YM}

In this section, we use the transmutation operator introduced previously to reconstruct known soft factors of YM amplitudes at leading and sub-leading orders, and prove the nonexistence of higher order soft factor which satisfies universality.
To avoid the treatment for complexity induced by momentum conservation, in this and next sections, we use the momentum conservation to eliminate all $k_n$ in each amplitude.

\subsection{Constraints on soft factors}
\label{subsec-softfac}

Consider the soft behavior of any $n$-point amplitude ${\cal A}_n$, with $k_i\to\tau k_i$, $\tau\to 0$. One can always expand the full amplitude as
in \eref{expan-general},
to acquire the formal factorization behavior ${\cal A}^{(a)_i}_n=S^{(a)_i}\,{\cal A}_{n-1}$ at any order. Since such formula is not the physically expected factorization, we need to impose appropriate constraints on soft factors.

The first important constraint is the universality, i.e., the expression of the soft factor is independent on the number of external legs. For $4$-point amplitudes, all allowed propagators have the form $1/s_{ij}$ where $i$ is the soft particle under consideration, thus the denominator of each term behaves as $\tau s_{ij}$ in the soft limit. Therefore, for the $4$-point case, the formula of soft factor at $a^{\rm th}$ order is restricted to
\bea
S^{(a)_i}_g=\sum_{j}\,{N^{(a)}_{ij}\over s_{ij}}\,,~~\label{softfac-condi1}
\eea
where the summation is over all $j$ those $1/s_{ij}$ contribute to ${\cal A}_n$. Then the universality requires the formula of soft factors in \eref{softfac-condi1} should also be satisfied for higher-point cases. It is impossible to extended the formula \eref{softfac-condi1} to incorporate propagators like $1/s_{ijk}$ from multi-particle channels without breaking the universality, since these propagators do not proportional to $1/\tau$ thus can not be on an equal footing with $1/s_{ij}$. If the formula \eref{softfac-condi1} for arbitrary number of external particles does not hold at the $a^{\rm th}$ order, then we say the universal soft factor does not exist at this order, since the independence on the number of external legs is violated.

For the YM case under consideration in this section, the numerator $N^{(a)}_{ij}(\epsilon_i,k_i,k_\ell)$ depends on $\epsilon_i$, $k_i$ for the soft gluon, and $k_\ell$ carried by other external gluons, but is independent of any $\epsilon_\ell$ with $\ell\neq i$, since both ${\cal A}^{(a)_i}_{\rm YM}(\vec{\pmb\sigma}_n)$ and ${\cal A}_{\rm YM}(\vec{\pmb\sigma}_n\setminus i)$ are linear on each $\epsilon_\ell$. We allow each soft factor to be an operator, acts on $(n-1)$-point amplitudes, then the independence on $\epsilon_\ell$ should be understood as that the effect of acting operator $N^{(a)}_{ij}(\epsilon_i,k_i,k_\ell)$ doe not break the linearity on $\epsilon_\ell$.

Furthermore, for the gluons of YM theory, we can extended the universality of soft behavior to a stronger version: the soft factors of pure YM amplitudes also holds for Yang-Mills-scalar (YMS) amplitudes in which gluons interact with BAS scalars. The reason is, the $3$-point YMS amplitude with two external scalars $1,3$ and one external gluon $2$, is related to the $3$-point YM amplitude via the operator $\partial_{\epsilon_1\cdot\epsilon_3}$ introduced in section \ref{subsec-operator}. The differential $\partial_{\epsilon_1\cdot\epsilon_3}$ is equivalent to the dimensional reduction, namely, consider the $(d+1)$-dimensional space-time, let the nonzero components of $\epsilon_1$ and $\epsilon_3$ to be in the extra dimension, and keep $\epsilon_2$ and all $k_p$ with $p\in\{1,2,3\}$ to be in the ordinary $d$-dimensional space-time. From $d$-dimensional point of view, $1$ and $3$ behaves as two scalars while $2$ behaves as a gluon. However, from the $(d+1)$-dimensional perspective, the $d$-dimensional scalar-scalar-gluon and gluon-gluon-gluon vertices are exactly the same interaction. Based on the above reason, it is nature to generalize the universality of gluon soft behaviors to the YMS case. Such extended universality will be useful in subsequent works.

Now we discuss the constraints from gauge invariance. To make the discussion clear, let us define the Ward identity operator as
\bea
{\cal W}_q\equiv\sum_{v}\,(k_q\cdot v)\,\partial_{\epsilon_q\cdot v}\,,
\eea
where the summation is over all Lorentz vectors $v^\mu$. Each numerator $N^{(a)}_{ij}(\epsilon_i,k_i,k_\ell)$ in \eref{softfac-condi1} should be consistent with gauge invariance for polarizations carried by external gluons, i.e.,
\bea
0=S^{(a)_i}_g\,{\cal W}_q\,{\cal A}_{\rm YM}(\vec{\pmb\sigma}_n\setminus i)={\cal W}_q\,S^{(a)_i}_g\,{\cal A}_{\rm YM}(\vec{\pmb\sigma}_n\setminus i)\,.~~\label{softfac-condi2}
\eea
In the above, the first equality holds because the $(n-1)$-point amplitude ${\cal A}_{\rm YM}(\vec{\pmb\sigma}_n\setminus i)$ in independent on $\epsilon_i$ when $q=i$, and states the gauge invariance of the amplitude ${\cal A}_{\rm YM}(\vec{\pmb\sigma}_n\setminus i)$ when $q\neq i$. The second is based on the gauge invariance of ${\cal A}_{\rm YM}(\vec{\pmb\sigma}_n)$ and the definition ${\cal A}^{(a)_i}_{\rm YM}(\vec{\pmb\sigma}_n)=S^{(a)_i}_g\,{\cal A}_{\rm YM}(\vec{\pmb\sigma}_n\setminus i)$. Thus, we conclude the commutativity $[{\cal W}_q,S^{(a)_i}_g]=0$, due to the relation in \eref{softfac-condi2}.

Such commutativity requires the soft operator $S^{(a)_i}_g$ to have the form
\bea
S^{(a)_i}_g=\sum_{j}\,\sum_{r=1}^{k}\,{P^{(a)}_{ij;r}(\epsilon_i,k_i,k_\ell)\over s_{ij}}\,{\cal O}^{(a)}_{ij;r}\,,~~\label{softfac-condi}
\eea
where each $P^{(a)}_{ij;r}(\epsilon_i,k_i,k_\ell)$ is a polynomial of Lorentz invariants, while each ${\cal O}^{(a)}_{ij;r}$ is an operator. The summation over integers $r$ means we allow more than one operators contribute to the numerator $N^{(a)}_{ij}(\epsilon_i,k_i,k_\ell)$, albeit all cases which will be encountered in this note are those $k=1$. The operator ${\cal O}^{(a)}_{ij;r}$ satisfies that when acting on any Lorentz invariant $k_p\cdot v$, the linearity on $k_p$ is kept, due to the following reason. The operator $S^{(a)_i}_g$ transmutes the Lorentz invariant $\epsilon_p\cdot v$ with $p\neq i$ as follows
\bea
S^{(a)_i}_g\,\Big(\epsilon_p\cdot v\Big)=\sum_j\,\sum_{r=1}^{k}\,{P^{(a)}_{ij;r}(\epsilon_i,k_i,k_\ell)\over s_{ij}}\,\epsilon_p\cdot v'_{j;r}\,,
\eea
without breaking the linearity on $\epsilon_p$, as discussed previously. Here we have replaced the Lorentz vector $v^\mu$ by new ones $(v'_{j;r})^\mu$, to reflect the action of ${\cal O}^{(a)}_{ij;r}$. Therefore, we have
\bea
{\cal W}_p\,S^{(a)_i}_g\,\Big(\epsilon_p\cdot v\Big)=\sum_j\,\sum_{r=1}^{k}\,{P^{(a)}_{ij}(\epsilon_i,k_i,k_\ell)\over s_{ij}}\,k_p\cdot v'_{j;r}\,.
\eea
Then the commutation relation $[{\cal W}_p,S^{(a)_i}_g]=0$ leads to
\bea
S^{(a)_i}_g\,\Big(k_p\cdot v\Big)=S^{(a)_i}_g\,{\cal W}_p\,\Big(\epsilon_p\cdot v\Big)=\sum_j\,\sum_{r=1}^{k}\,{P^{(a)}_{ij}(\epsilon_i,k_i,k_\ell)\over s_{ij}}\,k_p\cdot v'_{j;r}\,,
\eea
which means the operators ${\cal O}^{(a)}_{ij;r}$ do not affect the linearity on $k_p$.

With the expected formula \eref{softfac-condi} and the property of ${\cal O}^{(a)}_{ij;r}$ displayed above, now we are ready to study the existence of soft factors at each order.

\subsection{Leading order}
\label{subsec-leadingYM}

In this subsection, we use transmutation operators to investigate whether the leading order soft behavior of the YM amplitude ${\cal A}_{\rm YM}(\vec{\pmb\sigma}_n)$ can be represented as the factorized formula
\bea
{\cal A}^{(0)_i}_{\rm YM}(\vec{\pmb\sigma}_n)=S^{(0)_i}_g\,{\cal A}_{\rm YM}(\vec{\pmb\sigma}_n\setminus i)\,,~~\label{YM-fac-leading}
\eea
where the leading soft factor $S^{(0)_i}_g$ satisfies the required form in \eref{softfac-condi}.

We chose the combinatorial transmutation operator ${\cal T}[1,\cdots,n]$ to be
\bea
{\cal T}_0[1,\cdots,n]&=&\Big(\partial_{\epsilon_i\cdot k_{i-1}}-\partial_{\epsilon_i\cdot k_{i+1}}\Big)\,\Big(\prod_{a=i+2}^{n-1}\,\partial_{\epsilon_a\cdot k_{a-1}}\Big)\,\partial_{\epsilon_{i+1}\cdot k_{i-1}}\,\Big(\prod_{a=2}^{i-1}\,\partial_{\epsilon_a\cdot k_{a-1}}\Big)\,\partial_{\epsilon_1\cdot\epsilon_n}\,,~~\label{T0-np}
\eea
which creates the ordering $1,\cdots,n$ as follows:
\begin{itemize}
 \item Generating two endpoints $1$ and $n$ of the ordering.
 \item Inserting legs $a$ with $a\in\{2,\cdots,i-1\}$ between $1$ and $n$.
 \item Inserting the leg $(i+1)$ between $(i-1)$ and $n$.
 \item Inserting legs $a$ with $a\in\{i+2,\cdots,n-1\}$ between $(i+1)$ and $n$.
 \item Inserting the leg $i$ between $(i-1)$ and $(i+1)$.
\end{itemize}
In the operator \eref{T0-np}, all $\partial_{\epsilon_a\cdot k_n}$ in ${\cal I}_{(a-1)an}$ are removed, since all $k_n$ in the amplitude are eliminated by using momentum conservation. It is straightforward to recognize that
\bea
{\cal T}_0[1,\cdots,n]={\cal I}_{(i-1)i(i+1)}\,{\cal T}[1,\cdots,i-1,i+1,\cdots,n]\,,~~\label{T0-n-1p}
\eea
where the operator ${\cal T}[1,\cdots,i-1,i+1,\cdots,n]$ transmutes the $n$-point YM amplitude to YMS one as follows
\bea
{\cal T}[1,\cdots,i-1,i+1,\cdots,n]\,{\cal A}_{\rm YM}(\vec{\pmb\sigma}_n)={\cal A}_{\rm YMS}(1,\cdots,i-1,i+1,\cdots,n;i|\vec{\pmb\sigma}_n)\,.~~\label{YM-YMS}
\eea

Now we use the operator chosen in \eref{T0-np} to link the soft behaviors of gluons and BAS scalars together. We can expand the BAS and YM amplitudes by $\tau$, then the transmutation relation in \eref{trans-rela} reads
\bea
& &{1\over\tau}\,{\cal A}^{(0)_i}_{\rm BAS}(1,\cdots,n|\vec{\pmb\sigma}_n)+{\cal A}^{(1)_i}_{\rm BAS}(1,\cdots,n|\vec{\pmb\sigma}_n)+\tau\,{\cal A}^{(2)_i}_{\rm BAS}(1,\cdots,n|\vec{\pmb\sigma}_n)+\cdots\nn
&=&{\cal T}_0[1,\cdots,n]\,\Big({1\over\tau}\,{\cal A}^{(0)_i}_{\rm YM}(\vec{\pmb\sigma}_n)+{\cal A}^{(1)_i}_{\rm YM}(\vec{\pmb\sigma}_n)+\tau\,{\cal A}^{(2)_i}_{\rm YM}(\vec{\pmb\sigma}_n)+\cdots\Big)\,.
\eea
Since the operator ${\cal T}_0[1,\cdots,n]$ in \eref{T0-np} does not include any $\partial_{\epsilon_p\cdot k_i}$, it is independent of the soft parameter $\tau$. Consequently, we have
\bea
{\cal A}^{(a)_i}_{\rm BAS}(1,\cdots,n|\vec{\pmb\sigma}_n)={\cal T}_0[1,\cdots,n]\,{\cal A}^{(a)_i}_{\rm YM}(\vec{\pmb\sigma}_n)\,,~~\label{equ-T0}
\eea
holds at any order.

At the leading order, one can substitute the soft behavior of BAS amplitude given in \eref{for-soft-fac-s} and \eref{soft-fac-s-0}, to obtain
\bea
{\cal T}_0[1,\cdots,n]\,{\cal A}^{(0)_i}_{\rm YM}(\vec{\pmb\sigma}_n)&=&\Big({\delta_{(i-1)i}\over s_{(i-1)i}}+{\delta_{i(i+1)}\over s_{i(i+1)}}\Big)\,{\cal A}_{\rm BAS}(1,\cdots,i-1,i+1,\cdots,n|\vec{\pmb\sigma}_n\setminus i)\,.~~\label{p1-T0}
\eea
Suppose ${\cal A}^{(0)_i}_{\rm YM}(\vec{\pmb\sigma}_n)$ satisfies the factorized formula \eref{YM-fac-leading}, then we have
\bea
& &{\cal T}_0[1,\cdots,n]\,{\cal A}^{(0)_i}_{\rm YM}(\vec{\pmb\sigma}_n)\nn
&=&{\cal I}_{(i-1)i(i+1)}\,{\cal T}[1,\cdots,i-1,i+1,\cdots,n]\,{\cal A}^{(0)_i}_{\rm YM}(\vec{\pmb\sigma}_n)\nn
&=&{\cal I}_{(i-1)i(i+1)}\,{\cal A}^{(0)_i}_{\rm YMS}(1,\cdots,i-1,i+1,\cdots,n;i|\vec{\pmb\sigma}_n)\nn
&=&{\cal I}_{(i-1)i(i+1)}\,S^{(0)_i}_g\,{\cal A}_{\rm BAS}(1,\cdots,i-1,i+1,\cdots,n|\vec{\pmb\sigma}_n\setminus i)\,,~~\label{besub}
\eea
where the second equality uses the observation
\bea
{\cal T}[1,\cdots,i-1,i+1,\cdots,n]\,{\cal A}^{(0)_i}_{\rm YM}(\vec{\pmb\sigma}_n)={\cal A}^{(0)_i}_{\rm YMS}(1,\cdots,i-1,i+1,\cdots,n;i|\vec{\pmb\sigma}_n)\,,
\eea
based on the transmutation relation \eref{YM-YMS}.
The third uses
\bea
{\cal A}^{(0)_i}_{\rm YMS}(1,\cdots,i-1,i+1,\cdots,n;i|\vec{\pmb\sigma}_n)=S^{(0)_i}_g\,{\cal A}_{\rm BAS}(1,\cdots,i-1,i+1,\cdots,n|\vec{\pmb\sigma}_n\setminus i)\,,~~\label{softYMS}
\eea
which is indicated by the universality of soft behavior, namely, the formula \eref{softYMS} holds as long as \eref{YM-fac-leading} holds, with exactly the same soft factor.
Substituting \eref{besub} into \eref{p1-T0}, we find the equations
\bea
{\cal I}_{(i-1)i(i+1)}\,S^{(0)_i}_g
&=&{\delta_{(i-1)i}\over s_{(i-1)i}}+{\delta_{i(i+1)}\over s_{i(i+1)}}\,,~~\label{equ-T0-2}
\eea
hold for any $i\in\{2,\cdots,n-1\}$ (for $i=n-1$, the operator $\partial_{\epsilon_i\cdot k_{i+1}}$ in ${\cal I}_{(i-1)i(i+1)}$ should be removed, since we have eliminated all $k_n$ in the amplitude via momentum conservation).

The unique solution to equations \eref{equ-T0-2} for all $i\in\{2,\cdots,n-1\}$ is found to be
\bea
S^{(0)_i}_g=\sum_{j\neq i}\,{\delta_{ji}\,(\epsilon_i\cdot k_j)\over s_{ij}}\,.~~\label{YMs0-solu}
\eea
Notice that for $i=n-1$, the corresponding equation
\bea
\partial_{\epsilon_{n-1}\cdot k_{n-2}}\,S^{(0)_i}_g
&=&{\delta_{(n-2)(n-1)}\over s_{(n-2)(n-1)}}+{\delta_{(n-1)n}\over s_{(n-1)n}}
\eea
is not sufficient to determine the term $\delta_{ni}(\epsilon_i\cdot k_n)/s_{in}$ in the solution \eref{YMs0-solu}. This term is fixed by considering the gauge invariance for the polarization $\epsilon_i$. Start from the the ansatz
\bea
S^{(0)_i}_g={X\over s_{in}}+\sum_{j\neq i,n}\,{\delta_{ji}\,(\epsilon_i\cdot k_j)\over s_{ij}}\,,
\eea
the gauge invariance indicates
\bea
0={X\big|_{\epsilon_i\to k_i}\over s_{in}}+\sum_{j\neq i,n}\,{\delta_{ji}\over 2}\,,
\eea
then the relation $\sum_{k\neq i}\delta_{ki}=0$ together with the linearity on $\epsilon_i$ fix $X$ to be $\delta_{ni}(\epsilon_i\cdot n)$.
Comparing with the expected formula \eref{softfac-condi}, we see that the polynomial is $P^{(0)}_{ij;1}(\epsilon_i,k_i,k_{\ell})=\delta_{ji}(\epsilon_i\cdot k_j)$, the operator is the identity operator ${\cal O}^{(0)}_{ij;1}=\pmb1$. The symbol $\delta_{ji}$ requires the effective legs $j$ to be those adjacent to $i$ in the ordering $\vec{\pmb\sigma}_n$, thus ensured that the summation is for all $j$ those the corresponding $1/s_{ij}$ contribute to the amplitude.

It is worth to point out that the result in \eref{besub} also implies the commutativity
\bea
\Big[S^{(0)_i}_g,{\cal T}[1,\cdots,i-1,i+1,\cdots,n]\Big]=0\,,~~\label{commu}
\eea
due to the soft behavior \eref{YM-fac-leading} and the transmutation relation
\bea
{\cal T}[1,\cdots,i-1,i+1,\cdots,n]\,{\cal A}_{\rm YM}(\vec{\pmb\sigma}_n\setminus i)={\cal A}_{\rm BAS}(1,\cdots,i-1,i+1,\cdots,n|\vec{\pmb\sigma}_n\setminus i)\,.~~\label{transYM-n-1p}
\eea
In general, the above interpretation is not correct for the operator ${\cal T}[1,\cdots,i-1,i+1,\cdots,n]$ defined in \eref{T0-n-1p}, since momenta carried by gluons in the set $\{1,\cdots,n\}\setminus i$ violate momentum conservation. However, such interpretation makes sense in the soft limit $\tau\to0$. The commutativity in \eref{commu} can be generalized to arbitrary order as
\bea
& &{\cal T}[1,\cdots,i-1,i+1,\cdots,n]\,S^{(a)_i}_g\,{\cal A}_{\rm YM}(\vec{\pmb\sigma}_n\setminus i)\nn
&=&{\cal T}[1,\cdots,i-1,i+1,\cdots,n]\,{\cal A}^{(a)_i}_{\rm YM}(\vec{\pmb\sigma}_n)\nn
&=&{\cal A}^{(a)_i}_{\rm YMS}(1,\cdots,i-1,i+1,\cdots,n;i|\vec{\pmb\sigma}_n)\nn
&=&S^{(a)_i}_g\,{\cal A}_{\rm BAS}(1,\cdots,i-1,i+1,\cdots,n|\vec{\pmb\sigma}_n\setminus i)\nn
&=&S^{(a)_i}_g\,{\cal T}[1,\cdots,i-1,i+1,\cdots,n]\,{\cal A}_{\rm YM}(\vec{\pmb\sigma}_n\setminus i)\,,
\eea
if the soft factor satisfies the requirement in \eref{softfac-condi} exisit at the $a^{\rm th}$ order. The general commutation relation
\bea
\Big[S^{(a)_i}_g,{\cal T}[1,\cdots,i-1,i+1,\cdots,n]\Big]=0\,,~~\label{commu-gener}
\eea
will be useful in subsequent subsections.

\subsection{Sub-leading order}
\label{subsec-subleadingYM}

In this subsection, we continue to study the soft behavior of YM amplitudes at the sub-leading order. The relation \eref{equ-T0} also links the soft behaviors of YM and BAS amplitudes at the sub-leading order. However, since the factorized formula for BAS amplitudes is lacked, one can not repeat the manipulation in the previous subsection \ref{subsec-leadingYM}, to solve the sub-leading YM soft factor from the BAS one. Therefore, the operator ${\cal T}_0[1,\cdots,n]$ chosen in \eref{T0-np} and the relation \eref{equ-T0} are not effective for the current case. The above obstacle motivates us to chose new operator ${\cal T}_1[1,\cdots,n]$ which connects the sub-leading term of the YM side and the leading term of the BAS side together.

The new operator ${\cal T}_1[1,\cdots,n]$ is chosen to be that in \eref{definT-1}. Based on the assumption that all $k_n$ in amplitudes are removed via momentum conservation, we can remove all $\partial_{\epsilon_p\cdot k_n}$ in the insertions operators ${\cal I}_{(a-1)an}$ to obtain
\bea
{\cal T}_1[1,\cdots,n]=\Big(\prod_{a=2}^{n-1}\,\partial_{\epsilon_a\cdot k_{a-1}}\Big)\,\partial_{\epsilon_1\cdot\epsilon_n}\,.~~\label{T1-np}
\eea
In the soft limit, the operator \eref{T1-np} transmutes the expanded YM amplitude to expanded BAS amplitude as follows
\bea
& &{1\over\tau}\,{\cal A}^{(0)_i}_{\rm BAS}(1,\cdots,n|\vec{\pmb\sigma}_n)+{\cal A}^{(1)_i}_{\rm BAS}(1,\cdots,n|\vec{\pmb\sigma}_n)+\tau\,{\cal A}^{(2)_i}_{\rm BAS}(1,\cdots,n|\vec{\pmb\sigma}_n)+\cdots\nn
&=&{\cal T}_1[1,\cdots,n]\,\Big({1\over\tau}\,{\cal A}^{(0)_i}_{\rm YM}(\vec{\pmb\sigma}_n)+{\cal A}^{(1)_i}_{\rm YM}(\vec{\pmb\sigma}_n)+\tau\,{\cal A}^{(2)_i}_{\rm YM}(\vec{\pmb\sigma}_n)+\cdots\Big)\,.
\eea
Using the leading soft factor \eref{YMs0-solu}, it is direct to see that the operator ${\cal T}_1[1,\cdots,n]$ annihilates the leading order YM term
${\cal A}^{(0)_i}_{\rm YM}(\vec{\pmb\sigma}_n)$, since ${\cal T}_1[1,\cdots,n]$ involves the differential operator $\partial_{\epsilon_{i+1}\cdot k_i}$, while ${\cal A}^{(0)_i}_{\rm YM}(\vec{\pmb\sigma}_n)$ is independent of $k_i$. The operator $\partial_{\epsilon_{i+1}\cdot k_i}$ carries the parameter $1/\tau$ when $k_i\to\tau k_i$, therefore, the operator ${\cal T}_1[1,\cdots,n]$ transmutes the sub-leading YM term ${\cal A}^{(1)_i}_{\rm YM}(\vec{\pmb\sigma}_n)$ to the leading BAS term ${\cal A}^{(0)_i}_{\rm BAS}(1,\cdots,n|\vec{\pmb\sigma}_n)$. Thus, suppose the sub-leading soft behavior of YM amplitude satisfies the factorization
\bea
{\cal A}^{(1)_i}_{\rm YM}(\vec{\pmb\sigma}_n)=S^{(1)_i}_g\,{\cal A}_{\rm YM}(\vec{\pmb\sigma}_n\setminus i)\,,~~\label{YM-fac-sublead}
\eea
one can use the transmutation relation based on ${\cal T}_1[1,\cdots,n]$, to solve the soft factor $S^{(1)_i}_g$ from the leading soft behavior of BAS amplitude.

Based on above discussions, we can find the following relation for the assumed sub-leading soft factor $S^{(1)_i}_g$,
\bea
{\cal A}^{(0)_i}_{\rm BAS}(1,\cdots,n|\vec{\pmb\sigma}_n)&=&{\cal T}_1[1,\cdots,n]\,{\cal A}^{(1)_i}_{\rm YM}(\vec{\pmb\sigma}_n)\nn
&=&{\cal T}_1[1,\cdots,n]\,S^{(1)_i}_g\,{\cal A}_{\rm YM}(\vec{\pmb\sigma}_n\setminus i)\nn
&=&\partial_{\epsilon_{i+1}\cdot k_i}\,\partial_{\epsilon_i\cdot k_{i-1}}\,S^{(1)_i}_g\,{\cal P}_1\,{\cal A}_{\rm YM}(\vec{\pmb\sigma}_n\setminus i)\,,~~\label{BAS-0}
\eea
where the commutation relation
\bea
\Big[S^{(1)_i}_g,{\cal P}_1\Big]=0~~\label{commu-subleading}
\eea
with the operator ${\cal P}_1$ defined as
\bea
{\cal P}_1=\Big(\prod_{a=2}^{i-1}\,\partial_{\epsilon_a\cdot k_{a-1}}\Big)\,\Big(\prod_{a=i+2}^{n-1}\,\partial_{\epsilon_a\cdot k_{a-1}}\Big)\,\partial_{\epsilon_1\cdot\epsilon_n}\,,
\eea
is ensured by the commutativity in \eref{commu-gener}, since ${\cal P}_1$ is a subpart involved in ${\cal T}[1,\cdots,i-1,i+1,\cdots,n]$.
Substituting the leading soft factor of BAS amplitude \eref{soft-fac-s-0} and the transmutation relation \eref{transYM-n-1p} into \eref{BAS-0}, we arrive at the equations for $S^{(1)_i}_g$,
\bea
\partial_{\epsilon_{i+1}\cdot k_i}\,\partial_{\epsilon_i\cdot k_{i-1}}\,S^{(1)_i}_g\,{\cal P}_1\,{\cal A}_{\rm YM}(\vec{\pmb\sigma}_n\setminus i)=\Big({\delta_{(i-1)i}\over s_{(i-1)i}}+{\delta_{i(i+1)}\over s_{i(i+1)}}\Big)\,\partial_{\epsilon_{i+1}\cdot k_{i-1}}\,{\cal P}_1\,{\cal A}_{\rm YM}(\vec{\pmb\sigma}_n\setminus i)\,,~~\label{equ-T1}
\eea
hold for any $i\in\{2,\cdots,n-1\}$.  Notice that ${\cal P}_1\,{\cal A}_{\rm YM}(\vec{\pmb\sigma}_n)$ involves only one polarization $\epsilon_{i+1}$,
thus the above equations are convenient for analysing the effect of operator $S^{(1)_i}_g$.

To solve equations \eref{equ-T1}, we observe that the effect of differential $\partial_{\epsilon_{i+1}\cdot k_{i-1}}$ is turning $\epsilon_{i+1}\cdot k_{i-1}$ to $1$ and annihilating all terms without $\epsilon_{i+1}\cdot k_{i-1}$, due to the linear dependence on polarization $\epsilon_{i+1}$ of each physical amplitude. Similarly, the operator $\partial_{\epsilon_{i+1}\cdot k_i}\,\partial_{\epsilon_i\cdot k_{i-1}}$ turns $(\epsilon_{i+1}\cdot k_i)(\epsilon_i\cdot k_{i-1})$ to $1$ and annihilates all terms do not contain $(\epsilon_{i+1}\cdot k_i)(\epsilon_i\cdot k_{i-1})$. The Lorentz invariant $(\epsilon_{i+1}\cdot k_i)(\epsilon_i\cdot k_{i-1})$ under the action of $\partial_{\epsilon_{i+1}\cdot k_i}\,\partial_{\epsilon_i\cdot k_{i-1}}$ must be created by acting $S^{(1)_i}_g$ on ${\cal P}_1\,{\cal A}{\rm YM}(\vec{\pmb\sigma}_n\setminus i)$,
since ${\cal P}_1\,{\cal A}{\rm YM}(\vec{\pmb\sigma}_n\setminus i)$ is independent of $\epsilon_i$ and $k_i$. Thus, the operator $S^{(1)_i}_g$ should turn $\epsilon_{i+1}\cdot k_{i-1}$ to a new Lorentz invariant which involves $(\epsilon_{i+1}\cdot k_i)(\epsilon_i\cdot k_{i-1})$, namely,
\bea
S^{(1)_i}_g\,\epsilon_{i+1}\cdot k_{i-1}=\Big({\delta_{(i-1)i}\over s_{(i-1)i}}+{\delta_{i(i+1)}\over s_{i(i+1)}}\Big)\,(\epsilon_{i+1}\cdot k_i)(\epsilon_i\cdot k_{i-1})+C\,,
\eea
where $C$ is the potential part which is annihilated by $\partial_{\epsilon_{i+1}\cdot k_i}\partial_{\epsilon_i\cdot k_{i-1}}$. Using the gauge invariance requirement $\big[{\cal W}_i,S^{(1)_i}_g\big]=0$, one can fix the undetected part $C$, and obtain
\bea
S^{(1)_i}_g\,\epsilon_{i+1}\cdot k_{i-1}=\Big({\delta_{(i-1)i}\over s_{(i-1)i}}+{\delta_{i(i+1)}\over s_{i(i+1)}}\Big)\,(\epsilon_{i+1}\cdot f_i\cdot k_{i-1})\,,~~\label{YMs1-effe}
\eea
where the antisymmetric strength tensor is defined as $f_{a}^{\mu\nu}\equiv k^\mu_a\epsilon^\nu_a-\epsilon^\mu_a k^\nu_a$. Furthermore, the commutativity in\eref{commu-subleading} implies that the Lorentz invariants $\epsilon_1\cdot\epsilon_n$ and $\epsilon_a\cdot k_{a-1}$ with $a\in\{1,\cdots,i-1\}\cup\{i+2,\cdots,n-1\}$ are unaffected while $\epsilon_{i+1}\cdot k_{i-1}$ is transmuted as in \eref{YMs1-effe}. It means the operator $S^{(1)_i}_g$ should satisfy the Leibnitz rule. Consequently, the operator $S^{(1)_i}_g$ is found to be
\bea
S^{(1)_i}_g=
\sum_{j\neq i}\,{\delta_{ij}\over s_{ij}}\,\Big(k_j\cdot f_i\cdot \partial_{k_j}+\epsilon_j\cdot f_i\cdot \partial_{\epsilon_j}\Big)\,,~~\label{sg1}
\eea
which is equivalent to
\bea
S^{(1)_i}_g=
\sum_{j\neq i}\,{\delta_{ji}\,(\epsilon_i\cdot J_j\cdot k_i)\over s_{ij}}\,,~~\label{YMs1-solu}
\eea
where $J_j$ serves as the angular momentum carried by the external particle $j$.

Comparing \eref{YMs1-solu} with the desired formula in \eref{softfac-condi}, we see that the polynomial $P^{(1)}_{ij;1}(\epsilon_i,k_i,k_\ell)$ is trivially $\delta_{ji}$, while the operator is ${\cal O}^{(1)}_{ij;1}=\epsilon_i\cdot J_j\cdot k_i$. Again, the symbol $\delta_{ji}$ ensures that the effective summation is for external legs $j$, which contribute $1/s_{ij}$ to the amplitude.

\subsection{Higher order}
\label{subsec-higherYM}

The leading and sub-leading soft factors in \eref{YMs0-solu} and \eref{YMs1-solu} are standard soft factors of tree YM amplitudes, found in literatures \cite{Casali:2014xpa,Schwab:2014xua}. In this subsection, we argue that the sub-sub-leading soft factor $S^{(2)_i}_g$ satisfies the expectation in \eref{softfac-condi} does not exist.

Similar as in the previous subsection, our method is to choose an operator ${\cal T}_2[1,\cdots,n]$ which relates the sub-sub-leading YM term ${\cal A}_{\rm YM}^{(2)_i}(\vec{\pmb\sigma}_n)$ to the leading BAS term ${\cal A}^{(0)_i}_{\rm BAS}(1,\cdots,n|\vec{\pmb\sigma}_n)$, then try to solve $S^{(2)_i}_g$ from $S^{(0)_i}_s$. Such operator is chosen as
\bea
{\cal T}_2[1,\cdots,n]=\Big(\partial_{\epsilon_{i-1}\cdot k_{i-2}}-\partial_{\epsilon_{i-1}\cdot k_i}\Big)\,
\Big(\prod_{a=i+1}^{n-1}\,\partial_{\epsilon_a\cdot k_{a-1}}\Big)\,\partial_{\epsilon_i\cdot k_{i-2}}\,\Big(\prod_{a=2}^{i-2}\,\partial_{\epsilon_a\cdot k_{a-1}}\Big)\,\partial_{\epsilon_1\cdot\epsilon_n}\,,~~\label{T2-np}
\eea
which is similar to the operator ${\cal T}_0[1,\cdots,n]$ in \eref{T0-np}, but with $(i+1)$ replaced by $i$. The operator ${\cal T}_2[1,\cdots,n]$ in \eref{T2-np} can be separated as
\bea
{\cal T}_2[1,\cdots,n]={\cal T}_{21}[1,\cdots,n]+{\cal T}_{21}[1,\cdots,n]\,,
\eea
where
\bea
{\cal T}_{21}[1,\cdots,n]=
\Big(\prod_{a=i+1}^{n-1}\,\partial_{\epsilon_a\cdot k_{a-1}}\Big)\,\partial_{\epsilon_i\cdot k_{i-2}}\,\Big(\prod_{a=2}^{i-1}\,\partial_{\epsilon_a\cdot k_{a-1}}\Big)\,\partial_{\epsilon_1\cdot\epsilon_n}\,,~~\label{T21-np}
\eea
and
\bea
{\cal T}_{22}[1,\cdots,n]=-\partial_{\epsilon_{i-1}\cdot k_i}\,
\Big(\prod_{a=i+1}^{n-1}\,\partial_{\epsilon_a\cdot k_{a-1}}\Big)\,\partial_{\epsilon_i\cdot k_{i-2}}\,\Big(\prod_{a=2}^{i-2}\,\partial_{\epsilon_a\cdot k_{a-1}}\Big)\,\partial_{\epsilon_1\cdot\epsilon_n}\,.~~\label{T22-np}
\eea
It is straightforward to verify that the operator ${\cal T}_{21}[1,\cdots,n]$ annihilates ${\cal A}^{(0)_i}_{\rm YM}(\vec{\pmb\sigma}_n)$,
since ${\cal T}_{21}[1,\cdots,n]$ includes the differential $\partial_{\epsilon_{i+1}\cdot k_i}$ but ${\cal A}^{(0)_i}_{\rm YM}(\vec{\pmb\sigma}_n)$ is independent of $k_i$. Meanwhile, the operator ${\cal T}_{22}[1,\cdots,n]$ annihilates both ${\cal A}^{(0)_i}_{\rm YM}(\vec{\pmb\sigma}_n)$
and ${\cal A}^{(1)_i}_{\rm YM}(\vec{\pmb\sigma}_n)$, since differentials $\partial_{\epsilon_{i+1}\cdot k_i}$ and $\partial_{\epsilon_{i-1}\cdot k_i}$ in ${\cal T}_{22}[1,\cdots,n]$ requires the bilinearity on $k_i$, but ${\cal A}^{(0)_i}_{\rm YM}(\vec{\pmb\sigma}_n)$ is independent of $k_i$ and ${\cal A}^{(1)_i}_{\rm YM}(\vec{\pmb\sigma}_n)$ is linear on $k_i$. Two operators ${\cal T}_{21}[1,\cdots,n]$ carry scale parameters $1/\tau$
and $1/\tau^2$, arise from $\partial_{\epsilon_{i+1}\cdot k_i}$ and $\partial_{\epsilon_{i+1}\cdot k_i}\partial_{\epsilon_{i-1}\cdot k_i}$, respectively. Therefore, at the $\tau^{-1}$ order we have
\bea
{\cal A}^{(0)_i}_{\rm BAS}(1,\cdots,n|\vec{\pmb\sigma}_n)={\cal T}_{21}[1,\cdots,n]\,{\cal A}^{(1)_i}_{\rm YM}(\vec{\pmb\sigma}_n)
+{\cal T}_{22}[1,\cdots,n]\,{\cal A}^{(2)_i}_{\rm YM}(\vec{\pmb\sigma}_n)\,.~~\label{T2-sati}
\eea

By employing the sub-leading soft behavior of YM amplitudes in \eref{YM-fac-sublead} and \eref{YMs1-solu}, one can find
\bea
& &{\cal T}_{21}[1,\cdots,n]\,{\cal A}^{(1)_i}_{\rm YM}(\vec{\pmb\sigma}_n)\nn
&=&\partial_{\epsilon_{i+1}\cdot k_i}\,\partial_{\epsilon_i\cdot k_{i-2}}\,S^{(1)_i}_g\,\Big(\prod_{a=i+2}^{n-1}\,\partial_{\epsilon_a\cdot k_{a-1}}\Big)\,\Big(\prod_{a=2}^{i-1}\,\partial_{\epsilon_a\cdot k_{a-1}}\Big)\,\partial_{\epsilon_1\cdot\epsilon_n}\,
{\cal A}_{\rm YM}(\vec{\pmb\sigma}_n\setminus i)\nn
&=&\Big({\delta_{(i-2)i}\over s_{(i-2)i}}+{\delta_{i(i+1)}\over s_{i(i+1)}}\Big)\,\partial_{\epsilon_{i+1}\cdot k_{i-2}}\,\Big(\prod_{a=i+2}^{n-1}\,\partial_{\epsilon_a\cdot k_{a-1}}\Big)\,\Big(\prod_{a=2}^{i-1}\,\partial_{\epsilon_a\cdot k_{a-1}}\Big)\,\partial_{\epsilon_1\cdot\epsilon_n}\,
{\cal A}_{\rm YM}(\vec{\pmb\sigma}_n\setminus i)\,,~~\label{T12-resul}
\eea
where the first equality uses the commutation relation \eref{commu-gener}, and the second uses the following property
\bea
S^{(1)_i}_g\,(\epsilon_{i+1}\cdot k_{i-2})=\Big({\delta_{(i-2)i}\over s_{(i-2)i}}+{\delta_{i(i+1)}\over s_{i(i+1)}}\Big)\,
(\epsilon_{i+1}\cdot f_i\cdot k_{i-2})\,,
\eea
as can be directly verified by using the definition of $S^{(1)_i}_g$ in \eref{YMs1-solu} and \eref{sg1}. Meanwhile, we also have
\bea
& &{\cal T}_{22}[1,\cdots,n]\,{\cal A}^{(2)_i}_{\rm YM}(\vec{\pmb\sigma}_n)\nn
&=&-\partial_{\epsilon_{i+1}\cdot k_i}\,\partial_{\epsilon_i\cdot k_{i-2}}\,\partial_{\epsilon_{i-1}\cdot k_i}\,S^{(2)_i}_g\,
\Big(\prod_{a=i+2}^{n-1}\,\partial_{\epsilon_a\cdot k_{a-1}}\Big)\,\Big(\prod_{a=2}^{i-2}\,\partial_{\epsilon_a\cdot k_{a-1}}\Big)\,\partial_{\epsilon_1\cdot\epsilon_n}\,{\cal A}_{\rm YM}(\vec{\pmb\sigma}_n\setminus i)\,,~~\label{T22-resul}
\eea
if the desired $S^{(2)_i}_g$ exist. In the above derivation, the commutativity in \eref{commu-gener} is used again.

Combining results in \eref{T12-resul}, \eref{T22-resul} and the leading soft behavior of BAS amplitude together, we get the equations
\bea
& &\partial_{\epsilon_{i+1}\cdot k_i}\,\partial_{\epsilon_i\cdot k_{i-2}}\,\partial_{\epsilon_{i-1}\cdot k_i}\,S^{(2)_i}_g\,{\cal P}_2\,{\cal A}_{\rm YM}(\vec{\pmb\sigma}_n\setminus i)\nn
&=&\Big[\Big({\delta_{(i-2)i}\over s_{(i-2)i}}+{\delta_{i(i+1)}\over s_{i(i+1)}}\Big)\,\partial_{\epsilon_{i+1}\cdot k_{i-2}}\,\partial_{\epsilon_{i-1}\cdot k_{i-2}}-\Big({\delta_{(i-1)i}\over s_{(i-1)i}}+{\delta_{i(i+1)}\over s_{i(i+1)}}\Big)\,\partial_{\epsilon_{i+1}\cdot k_{i-1}}\,\partial_{\epsilon_{i-1}\cdot k_{i-2}}\Big]\,{\cal P}_2\,{\cal A}_{\rm YM}(\vec{\pmb\sigma}_n\setminus i)\,,~~\label{equ-T2}
\eea
hold for any $i\in\{2,\cdots,n-1\}$, with
\bea
{\cal P}_2=\Big(\prod_{a=i+2}^{n-1}\,\partial_{\epsilon_a\cdot k_{a-1}}\Big)\,\Big(\prod_{a=2}^{i-2}\,\partial_{\epsilon_a\cdot k_{a-1}}\Big)\,\partial_{\epsilon_1\cdot\epsilon_n}\,.
\eea
The above equations imply that the operator $S^{(2)_i}_g$ should transmute the Lorentz invariant $(\epsilon_{i+1}\cdot k_{i-2})(\epsilon_{i-1}\cdot k_{i-2})$ or $(\epsilon_{i+1}\cdot k_{i-1})(\epsilon_{i-1}\cdot k_{i-2})$ to a new Lorentz invariant, which contains a part proportional to $(\epsilon_{i+1}\cdot k_i)(\epsilon_i\cdot k_{i-2})(\epsilon_{i-1}\cdot k_i)$. For the first case, the bilinearity on $k_{i-2}$ is turned to the linearity. For the second case, the linearity on $k_{i-1}$ is turned to the independence.
Both two situations can not be realized via a polynomial $P^{(a)}_{ij}(\epsilon_i,k_i,k_\ell)$ and an operator ${\cal O}_{ij}$ which keeps the linearity on any $k_p$, as required in \eref{softfac-condi}. Thus, we conclude that the YM soft factor satisfies our expectation can not be found at the sub-sub-leading order.

In equations \eref{equ-T2}, we have used the transmutation relation
\bea
{\cal A}_{\rm BAS}(1,\cdots,i-1,i+1,\cdots,n|\vec{\pmb\sigma}_n\setminus i)=\partial_{\epsilon_{i+1}\cdot k_{i-1}}\,\partial_{\epsilon_{i-1}\cdot k_{i-2}}\,{\cal P}_2\,{\cal A}_{\rm YM}(\vec{\pmb\sigma}_n\setminus i)\,.~~\label{inser}
\eea
Instead of the above one, we can also consider the equivalent relation
\bea
{\cal A}_{\rm BAS}(1,\cdots,i-1,i+1,\cdots,n|\vec{\pmb\sigma}_n\setminus i)=\Big(\partial_{\epsilon_{i-1}\cdot k_{i-2}}-\partial_{\epsilon_{i-1}\cdot k_{i+1}}\Big)\,\partial_{\epsilon_{i+1}\cdot k_{i-2}}\,{\cal P}_2\,{\cal A}_{\rm YM}(\vec{\pmb\sigma}_n\setminus i)\,,~~\label{another-inser}
\eea
where the operator $\partial_{\epsilon_{i+1}\cdot k_{i-2}}$ serves as the insertion operator ${\cal I}_{(i-2)(i+1)n}$ which inserts the leg $(i+1)$
between $(i-2)$ and $n$, while the operator $(\partial_{\epsilon_{i-1}\cdot k_{i-2}}-\partial_{\epsilon_{i-1}\cdot k_{i+1}})$ is interpreted as ${\cal I}_{(i-2)(i-1)(i+1)}$, which inserts $(i-1)$ between $(i-2)$ and $(i+1)$. Replacing \eref{inser} by \eref{another-inser}, the equations \eref{equ-T2} are modified to
\bea
& &\partial_{\epsilon_{i+1}\cdot k_i}\,\partial_{\epsilon_i\cdot k_{i-2}}\,\partial_{\epsilon_{i-1}\cdot k_i}\,S^{(2)_i}_g\,{\cal P}_2\,{\cal A}_{\rm YM}(\vec{\pmb\sigma}_n\setminus i)\nn
&=&\Big[\Big({\delta_{(i-2)i}\over s_{(i-2)i}}+{\delta_{i(i+1)}\over s_{i(i+1)}}\Big)\,\partial_{\epsilon_{i+1}\cdot k_{i-2}}\,\partial_{\epsilon_{i-1}\cdot k_{i-2}}-\Big({\delta_{(i-1)i}\over s_{(i-1)i}}+{\delta_{i(i+1)}\over s_{i(i+1)}}\Big)\,\partial_{\epsilon_{i-1}\cdot k_{i-2}}\,\partial_{\epsilon_{i+1}\cdot k_{i-2}}\nn
& &+\Big({\delta_{(i-1)i}\over s_{(i-1)i}}+{\delta_{i(i+1)}\over s_{i(i+1)}}\Big)\,\partial_{\epsilon_{i-1}\cdot k_{i+1}}\,\partial_{\epsilon_{i+1}\cdot k_{i-2}}\Big]\,{\cal P}_2\,{\cal A}_{\rm YM}(\vec{\pmb\sigma}_n\setminus i)\,.~~\label{equ-T2-2}
\eea
For the above equations, the solution $S^{(2)_i}_g$ is also forbidden, since the similar analysis indicates the effect of turning the bilinearity on $k_{i-2}$ to the linearity, or turning the linearity on $k_{i+1}$ to the independence.

The similar argument can also be applied to exclude the solution of $S^{(a)_i}_g$ satisfying the requirement \eref{softfac-condi},
with $a\geq3$. For instance, one can choose the differentials those create the ordering $1,\cdots,i-3,i,\cdots,n$ first, then insert
$(i-2)$ between $(i-3)$ and $i$, and subsequently insert $(i-1)$ between $(i-2)$ and $i$. Such combinatorial operator can be decomposed
into three parts which carry scale parameters $1/\tau$, $1/\tau^2$ and $1/\tau^3$ respectively, therefore transmutes the summation of leading,
sub-leading and sub-sub-leading terms of YM amplitude to the leading contribution of BAS amplitude. Then one can observe the similar phenomenon,
the assumed soft operators $S^{(2)_i}_g$ and $S^{(3)_i}_g$ decrease the power of some external momenta thus are forbidden. It is easy to see that when
$k_i$ appears more than once in at lest one part of the combinatorial operator, then the above phenomenon, which excludes the solution under the constraint \eref{softfac-condi}, always happen.

\section{Soft behavior of GR amplitudes}
\label{sec-GR}

In this section, we study the factorization of GR amplitudes in the soft limit,
\bea
{\cal A}^{(a)_i}_{\rm GR}({\pmb h}_n)=S^{(a)_i}_h\,{\cal A}_{\rm GR}({\pmb h}_n\setminus i)\,,~~\label{GR-fac}
\eea
with assumed soft factors $S^{(a)_i}_h$, at $a^{\rm th}$ order.
Through the argument paralleled to that in section \ref{subsec-softfac}, we expect soft factors of GR amplitudes to take the form
\bea
S^{(a)_i}_h=\sum_{j}\,{N_{ij}\over s_{ij}}=\sum_{j}\,\sum_{r=1}^k\,{P^{(a)}_{ij;r}(\varepsilon_i,k_i,k_\ell)\over s_{ij}}\,{\cal O}^{(a)}_{ij;r}\,,~~\label{softfac-condi-GR}
\eea
where the operators ${\cal O}^{(a)}_{ij;r}$ maintain the linearity on $k_p$ when acting on $k_p\cdot v$, similar as its YM counterpart in \eref{softfac-condi}. Here $\varepsilon_i^{\mu\nu}$ is the polarization tensor carried by the soft graviton $i$. Meanwhile, the commutation relation
in \eref{commu-gener} is now extended to
\bea
\Big[S^{(a)_i}_h,\W{\cal T}[1,\cdots,i-1,i+1,\cdots,n]\Big]=0\,,~~\label{commu-gener-GR}
\eea

According to the transmutation in \eref{trans-rela}, the combinatorial operator $\W{\cal T}[1,\cdots,n]$ turns the GR amplitudes to the YM ones with specific ordering $1,\cdots,n$.
For such YM amplitudes ${\cal A}_{\rm YM}(1,\cdots,n)$ , the soft factors in \eref{YMs0-solu} and \eref{YMs1-solu}
are reduced to
\bea
S^{(0)_i}_g={\epsilon_i\cdot k_{i-1}\over s_{i(i-1)}}-{\epsilon_i\cdot k_{i+1}\over s_{i(i+1)}}\,,~~\label{YM-sfc-0}
\eea
and
\bea
S^{(1)_i}_g={\epsilon_i\cdot J_{i-1}\cdot k_i\over s_{i(i-1)}}-{\epsilon_i\cdot J_{i+1}\cdot k_i\over s_{i(i+1)}}\,.~~\label{YM-sfc-1}
\eea

We will find that the consistent sub-leading and sub-sub-leading soft factors in literatures \cite{Cachazo:2014fwa,Schwab:2014xua,Afkhami-Jeddi:2014fia,Zlotnikov:2014sva} should be defined for pure Einstein gravity. On the other hand, the transmutation operators make sense for extended gravity that Einstein gravity couples to $2$-form and dilaton field, whose amplitudes manifest the double copy structure. This gap complicates the discussion. Thus, it is worth to explain the double copy structure and its implications in more detail. We do this in subsection \ref{subsec-CHY} by employing the CHY formula. Then, in subsequent subsections, we rederive leading, sub-leading and sub-sub-leading soft factors for GR amplitudes,
and prove the nonexistence of higher order soft factor which satisfies the expectation \eref{softfac-condi-GR}.

\subsection{CHY formula and double copy structure}
\label{subsec-CHY}

The well known Cachazo-He-Yuan (CHY) formula manifests the double copy structure of GR amplitudes, which will play the important role in this section.
In CHY formula, the integrands for GR, YM, and BAS theories are \cite{Cachazo:2013hca,Cachazo:2013iea}
\bea
{\cal I}_{\rm GR}&=&{\rm Pf}'\Psi(\epsilon_p,k_p)\,{\rm Pf}'\W\Psi(\W\epsilon_p,k_p)\,,\nn
{\cal I}_{\rm YM}&=&{\rm Pf}'\Psi(\epsilon_p,k_p)\,{\rm PT}(\vec{\pmb\sigma}_n)\,,\nn
{\cal I}_{\rm BAS}&=&{\rm PT}(\vec{\pmb\sigma}_n)\,{\rm PT}(\vec{\pmb\sigma}'_n)\,.~~\label{CHY-integrand}
\eea
In the above, ${\rm Pf}'\Psi(\epsilon_p,k_p)$ encodes the reduced Pffafian of the matrix $\Psi(\epsilon_p,k_p)$ which depends
on external polarizations $\epsilon_p$ and momenta $k_p$, while ${\rm PT}(\vec{\pmb\sigma}_n)$ denotes the Parke-Taylor factor with the ordering
$\vec{\pmb\sigma}_n$ which is independent of any external kinematic variable. The polarization tensor carried by a graviton is decomposed as
$\varepsilon_p^{\mu\nu}=\epsilon_p^\mu\W\epsilon_p^\nu$. For the extended gravity, $\epsilon_p$
and $\W\epsilon_p$ are independent of each other. For Einstein gravity, $\epsilon_p$
and $\W\epsilon_p$ are equivalent. The tree amplitudes for above three theories can be obtained by doing the contour integration for integrands given in \eref{CHY-integrand}, with the poles
determined by so called scattering equations.

As demonstrated in \cite{Zhou:2018wvn,Bollmann:2018edb}, the transmutation operator ${\cal T}[\vec{\pmb\sigma}_n]$ transmutes ${\rm Pf}'\Psi(\epsilon_p,k_p)$ to ${\rm PT}(\vec{\pmb\sigma}_n)$, and analogously $\W {\cal T}[\vec{\pmb\sigma}'_n]$ transmutes ${\rm Pf}'\W\Psi(\W\epsilon_p,k_p)$ to ${\rm PT}(\vec{\pmb\sigma}'_n)$. In other words, they connect CHY integrands of GR, YM and BAS
theories together.

Each ${\rm Pf}'\Psi(\epsilon_p,k_p)$ (or ${\rm Pf}'\W\Psi(\W\epsilon_p,k_p)$) can be expanded to Parke-Taylor
factors as
\bea
{\rm Pf}'\Psi(\epsilon_p,k_p)=\sum_{\vec{\pmb\sigma}_n}\,C(\epsilon_p,k_p,\vec{\pmb\sigma}_n)\,{\rm PT}(\vec{\pmb\sigma}_n)\,,
\eea
where the coefficients $C(\epsilon_p,k_p,\vec{\pmb\sigma}_n)$ are polynomials of Lorentz invariants arise from external polarizations
and momenta. Consequently, the GR and YM amplitudes can be expanded to BAS amplitudes, namely,
\bea
{\cal A}_{\rm GR}(\pmb h_n)&=&\sum_{\vec{\pmb\sigma}_n}\,\sum_{\vec{\pmb\sigma}'_n}\,C(\epsilon_p,k_p,\vec{\pmb\sigma}_n)\,{\cal A}_{\rm BAS}(\vec{\pmb\sigma}_n|\vec{\pmb\sigma}'_n)\,\W C(\W\epsilon_p,k_p,\vec{\pmb\sigma}'_n)\,,\nn
{\cal A}_{\rm YM}(\vec{\pmb\sigma}_n)&=&\sum_{\vec{\pmb\sigma}'_n}\,C(\epsilon_p,k_p,\vec{\pmb\sigma}'_n)\,{\cal A}_{\rm BAS}(\vec{\pmb\sigma}_n|\vec{\pmb\sigma}'_n)\,.~~\label{expan}
\eea
This structure indicates that the transmutation operator ${\cal T}[1,\cdots,n]$ turns $C(\epsilon_p,k_p,\vec{\pmb\sigma}_n)$ with
$\vec{\pmb\sigma}_n=\{1,\cdots,n\}$ to $1$, and annihilates all other $C(\epsilon_p,k_p,\vec{\pmb\sigma}_n)$. The analogous statement
holds for $\W{\cal T}[1,\cdots,n]$ and $\W C(\W\epsilon_p,k_p,\vec{\pmb\sigma}_n)$.

The CHY integrands in \eref{CHY-integrand} and the expansions in \eref{expan} indicate that the Lorentz invariants in the form $\epsilon_p\cdot\W\epsilon_k$ never occur in amplitudes of extended gravity. In subsequent subsections, we will maintain such character carefully when
deriving soft operators.

Another new situation indicated by the double copy structure is as follows. Since transmutation operators act on only one of two reduced Pfaffians in CHY integrands \eref{CHY-integrand} (or equivalently one of two coefficients in expansions \eref{expan}), while another one also contributes to soft behaviors, when performing such operators to solve soft factors, new undetectable terms which can not be determined by imposing gauge invariance will occur. We will see the examples when considering the sub-leading and sub-sub-leading soft behaviors of GR amplitudes.

\subsection{Leading and sub-leading orders}
\label{subsec-leadingsubleading-GR}

In this subsection, we derive the soft factors of GR amplitudes at leading and sub-leading orders. The method in this subsection is similar as that in section \ref{subsec-leadingYM}. We choose the combinatorial operator $\W{\cal T}_0[1,\cdots,n]$ as
\bea
\W{\cal T}_0[1,\cdots,n]&=&\Big(\partial_{\W\epsilon_i\cdot k_{i-1}}-\partial_{\W\epsilon_i\cdot k_{i+1}}\Big)\,\Big(\prod_{a=i+2}^{n-1}\,\partial_{\W\epsilon_a\cdot k_{a-1}}\Big)\,\partial_{\W\epsilon_{i+1}\cdot k_{i-1}}\,\Big(\prod_{a=2}^{i-1}\,\partial_{\W\epsilon_a\cdot k_{a-1}}\Big)\,\partial_{\W\epsilon_1\cdot\W\epsilon_n}\nn
&=&\W{\cal I}_{(i-1)i(i+1)}\,\W{\cal T}[1,\cdots,i-1,i+1,\cdots,n]\,,~~\label{T0-gr}
\eea
which is paralleled to the definition of ${\cal T}_0[1,\cdots,n]$ in \eref{T0-np}, with each $\epsilon_p$ replaced by $\W\epsilon_p$. The operator
$\W{\cal T}_0[1,\cdots,n]$ connects soft behaviors of GR and YM amplitudes as
\bea
{\cal A}^{(a)_i}_{\rm YM}(1,\cdots,n)=\W{\cal T}_0[1,\cdots,n]\,{\cal A}^{(a)_i}_{\rm GR}({\pmb h}_n)\,,~~\label{equ-T0}
\eea
allows us to solve $S^{(0)_i}_h$ and a part of $S^{(1)_i}_h$ from $S^{(0)_i}_g$ and $S^{(1)_i}_g$, respectively.

At the leading order $a=0$, we have
\bea
{\cal A}^{(0)_i}_{\rm YM}(1,\cdots,n)&=&\W{\cal T}_0[1,\cdots,n]\,{\cal A}^{(0)_i}_{\rm GR}({\pmb h}_n)\nn
&=&\W{\cal I}_{(i-1)i(i+1)}\,S^{(0)_i}_h\,\W{\cal T}[1,\cdots,i-1,i+1,\cdots,n]\,{\cal A}_{\rm GR}({\pmb h}_n\setminus i)\nn
&=&\W{\cal I}_{(i-1)i(i+1)}\,S^{(0)_i}_h\,{\cal A}_{\rm YM}(1,\cdots,i-1,i+1,\cdots,n)\,,
\eea
where the second equality uses the leading order factorization \eref{GR-fac}, as well as the commutativity \eref{commu-gener-GR}. Substituting the leading soft factor of YM amplitude in \eref{YM-sfc-0}, we get the equations
\bea
\Big(\partial_{\W\epsilon_i\cdot k_{i-1}}-\partial_{\W\epsilon_i\cdot k_{i+1}}\Big)\,S^{(0)_i}_h={\epsilon_i\cdot k_{i-1}\over s_{i(i-1)}}-{\epsilon_i\cdot k_{i+1}\over s_{i(i+1)}}\,,~~\label{equ-T0-GR}
\eea
hold for any $i\in\{2,\cdots,n-1\}$. The solution to the above equations \eref{equ-T0-GR} is
\bea
S^{(0)_i}_h=\sum_{j\neq i}\,{(\epsilon_i\cdot k_j)\,(\W\epsilon_i\cdot k_j)\over s_{ij}}=\sum_{j\neq i}\,{k_j\cdot\varepsilon_i\cdot k_j\over s_{ij}}\,.~~\label{GRs0-solu}
\eea
coincides with the leading soft factor given in \cite{Cachazo:2014fwa,Schwab:2014xua,Afkhami-Jeddi:2014fia}.
The gauge invariance of the above soft factor $S^{(0)_i}_h$ is ensured by momentum conservation and on-shell condition. For instance, replacing $\epsilon_i$ by $k_i$ yields
\bea
\sum_{j\neq i}\,{\W\epsilon_i\cdot k_j\over 2}=-{\W\epsilon_i\cdot k_i\over2}=0\,.
\eea
The gauge invariance for polarization $\W\epsilon_i$ is analogous.

The paralleled procedure leads to equations at the sub-leading order
\bea
& &\Big(\partial_{\W\epsilon_i\cdot k_{i-1}}-\partial_{\W\epsilon_i\cdot k_{i+1}}\Big)\,S^{(1)_i}_h\,{\cal A}_{\rm YM}(1,\cdots,i-1,i+1,\cdots,n)\nn
&=&\Big({\epsilon_i\cdot J_{i-1}\cdot k_i\over s_{i(i-1)}}-{\epsilon_i\cdot J_{i+1}\cdot k_i\over s_{i(i+1)}}\Big)\,{\cal A}_{\rm YM}(1,\cdots,i-1,i+1,\cdots,n)\,,~~\label{equ-T1-GR}
\eea
where the soft factor in \eref{YM-sfc-1} is used. The solution to above equations \eref{equ-T1-GR} is found to be
\bea
S_1&=&\sum_{j\neq i}\,{(\epsilon_i\cdot {\cal J}_j\cdot k_i)\,(\W\epsilon_i\cdot k_j)\over s_{ij}}\,,~~\label{GRs1-solu1}
\eea
where ${\cal J}_j$ are operators
\bea
{\cal J}_j^{\mu\nu}\,k_j^\rho= k_j^{\mu}\,
{\partial k_j^\rho\over\partial k_{j,\nu}}-k_j^{\nu}\,{\partial k_j^\rho\over\partial k_{j,\mu}}\,,~~~~
{\cal J}_j^{\mu\nu}\,\epsilon_j^\rho=\big(\eta^{\nu\rho}\,\delta^\mu_\sigma-\eta^{\mu\rho}\,\delta^\nu_\sigma\big)\,
\epsilon^\sigma_j\,,~~\label{J-left}
\eea
which do not act on Lorentz invariants contributed by $\W C(\W\epsilon_p,k_p,\vec{\pmb\sigma}_n)$ in \eref{expan}.
When transmuted to YM amplitudes with only one reduced Pfaffian in the corresponding CHY integrand,
these operators with the effects in \eref{J-left} restore the standard angular momentum operators. For the extended gravity, they can not be
interpreted as angular momentum operators, since they do not act on all orbital and spin parts of an external graviton.
The reason for introducing the above ${\cal J}_j$ is to maintain the correct sub-leading soft factor for YM amplitudes and the double copy structure in \eref{CHY-integrand} and \eref{expan} simultaneously. Suppose we allow the operator ${\cal J}_j$ to act on $\W C(\W\epsilon_p,k_p,\vec{\pmb\sigma}_n)$, then $\epsilon_\ell\cdot\W\epsilon_k$ will occur.

Based on symmetry, it is natural to expect another block
\bea
S_2&=&\sum_{j\neq i}\,{(\epsilon_i\cdot k_j)\,(\W\epsilon_i\cdot \W{\cal J}_j\cdot k_i)\over s_{ij}}\,,~~\label{GRs1-solu2}
\eea
which do not act on Lorentz invariants from $C(\epsilon_p,k_p,\vec{\pmb\sigma}_n)$, and is connected to $S_1$ by exchanging $\epsilon$ and $\W\epsilon$.
This part can not be detected by the operator $\W{\cal T}_0[1,\cdots,n]$, and will be found in next subsection \ref{subsec-subsubleading-GR} by using the operator $\W{\cal T}_1[1,\cdots,n]$. The full soft factor at the sub-leading order is the summation of two parts $S_1$ and $S_2$, namely,
\bea
S^{(1)_i}_h=S_1+S_2=\sum_{j\neq i}\,{(\epsilon_i\cdot k_j)\,(\W\epsilon_i\cdot\W{\cal J}_j\cdot k_i)+(\W\epsilon_i\cdot k_j)\,(\epsilon_i\cdot {\cal J}_j\cdot k_i)\over s_{ij}}\,.~~\label{GRs1-solu}
\eea

In practice, the formula \eref{GRs1-solu} does not make sense, due to the following reason. By definition, the operators ${\cal J}_j$ do not act on $k_\ell\cdot k_k$ from $\W C(\W\epsilon_p,k_p,\vec{\pmb\sigma}_n)$, while operators $\W{\cal J}_j$ do not act on $k_\ell\cdot k_k$ from $C(\epsilon_p,k_p,\vec{\pmb\sigma}_n)$. However, it is impossible to distinguish the origins of these $k_\ell\cdot k_k$ in each amplitude.
The situation has changed dramatically if we restrict ourselves to standard Einstein gravity. For Einstein gravity, in which $\epsilon^\mu_p=\W\epsilon^\mu_p$, the solution \eref{GRs1-solu} is reduced to
\bea
S^{(1)_i}_h=\sum_{j\neq i}\,{(\epsilon_i\cdot k_j)\,(\epsilon_i\cdot J_j\cdot k_i)\over s_{ij}}
=\sum_{j\neq i}\,{k_j\cdot\varepsilon_i\cdot J_j\cdot k_i\over s_{ij}}\,,~~\label{GRs1-solu-Ein}
\eea
since in this case ${\rm Pf}'\Psi(\epsilon_p,k_p)$ and ${\rm Pf}'\W\Psi(\W\epsilon_p,k_p)$ are equivalent to each other. In \eref{GRs1-solu-Ein}, the operators $J_j$ are angular momentum operators which act on all $k_j$, $\epsilon_j$ and $\W\epsilon_j$, therefore affect orbital and spin parts of the $j^{\rm th}$ graviton in the correct manner.
The equivalence between ${\rm Pf}'\Psi(\epsilon_p,k_p)$ and ${\rm Pf}'\W\Psi(\W\epsilon_p,k_p)$ implies that it is not necessary to distinguish the origins of $k_\ell\cdot k_k$ for the current case.
The sub-leading soft factor in \eref{GRs1-solu-Ein} is also coincide with the result in \cite{Cachazo:2014fwa,Schwab:2014xua,Afkhami-Jeddi:2014fia}.

\subsection{Sub-leading and sub-sub-leading orders}
\label{subsec-subsubleading-GR}

In this subsection, we derive the $S_2$ part of the soft factor $S^{(1)_i}_h$, as well as a part of $S^{(2)_i}_h$, by using the operator
\bea
\W{\cal T}_1[1,\cdots,n]=\Big(\prod_{a=2}^{n-1}\,\partial_{\W\epsilon_a\cdot k_{a-1}}\Big)\,\partial_{\W\epsilon_1\cdot\W\epsilon_n}\,,~~\label{T1-gr}
\eea
obtained by replacing $\epsilon_p$ with $\W\epsilon_p$ in \eref{T1-np}. This operator annihilates the leading GR term ${\cal A}^{(0)_i}_{\rm GR}(\pmb{h}_n)$, and links GR terms at $(a+1)^{\rm th}$ order to YM terms at $a^{\rm th}$ order,
\bea
{\cal A}^{(a)_i}_{\rm YM}(1,\cdots,n)=\W{\cal T}_1[1,\cdots,n]\,{\cal A}^{(a+1)_i}_{\rm GR}(\pmb{h}_n)\,.
\eea
Such connection allows us to detect $S^{(a+1)_i}_h$ by substituting known $S^{(a)_i}_g$.

At the sub-leading order, we have
\bea
{\cal A}^{(0)_i}_{\rm YM}(1,\cdots,n)&=&\W{\cal T}_1[1,\cdots,n]\,{\cal A}^{(1)_i}_{\rm GR}(\pmb{h}_n)\nn
&=&\partial_{\W\epsilon_{i+1}\cdot k_i}\,\partial_{\W\epsilon_i\cdot k_{i-1}}\,S^{(1)_i}_h\,\Big[\Big(\prod_{a=2}^{i-1}\,\partial_{\W\epsilon_a\cdot k_{a-1}}\Big)\,\Big(\prod_{a=i+2}^{n-1}\,\partial_{\W\epsilon_a\cdot k_{a-1}}\Big)\,\partial_{\W\epsilon_1\cdot\W\epsilon_n}\Big]\,{\cal A}_{\rm GR}(\pmb {h}_n\setminus i)\,,~~\label{ste1}
\eea
where the commutativity in \eref{commu-gener-GR} is used again. Substituting the soft factor in \eref{YM-sfc-0}, as well as the transmutation relation
\bea
{\cal A}_{\rm YM}(1,\cdots,i-1,i+1,\cdots,n)=\W{\cal T}[1,\cdots,i-1,i+1,\cdots,n]\,{\cal A}_{\rm GR}(\pmb {h}_n\setminus i)\,,~~\label{GR-YM-n-1p}
\eea
into the relation \eref{ste1}, we find the equations
\bea
\partial_{\W\epsilon_{i+1}\cdot k_i}\,\partial_{\W\epsilon_i\cdot k_{i-1}}\,S^{(1)_i}_h\,\W{\cal P}_1\,{\cal A}_{\rm GR}({\pmb h}_n\setminus i)=\Big({\epsilon_i\cdot k_{i-1}\over s_{(i-1)i}}-{\epsilon_i\cdot k_{i+1}\over s_{i(i+1)}}\Big)\,\partial_{\W\epsilon_{i+1}\cdot k_{i-1}}\,\W{\cal P}_1\,{\cal A}_{\rm GR}({\pmb h}_n\setminus i)\,,~~\label{equ-T1-GR-2}
\eea
with
\bea
\W{\cal P}_1=\Big(\prod_{a=2}^{i-1}\,\partial_{\W\epsilon_a\cdot k_{a-1}}\Big)\,\Big(\prod_{a=i+2}^{n-1}\,\partial_{\W\epsilon_a\cdot k_{a-1}}\Big)\,\partial_{\W\epsilon_1\cdot\W\epsilon_n}\,.
\eea
Through the technic extremely similar to that for solving equation \eref{equ-T1}, we find the solution to equation \eref{equ-T1-GR-2} is $S_2$ in \eref{GRs1-solu2}. As discussed in the previous subsection \ref{subsec-leadingsubleading-GR}, the combination of two parts $S_1$ and $S_2$ leads to the full sub-leading soft factor, which can be reduced to the standard formula in Einstein gravity.

At the sub-sub-leading order, the same manipulation gives the similar equations
\bea
\partial_{\W\epsilon_{i+1}\cdot k_i}\,\partial_{\W\epsilon_i\cdot k_{i-1}}\,S^{(2)_i}_h\,\W{\cal P}_1\,{\cal A}_{\rm GR}({\pmb h}_n\setminus i)=\Big({\epsilon_i\cdot {\cal J}_{i-1}\cdot k_i\over s_{(i-1)i}}-{\epsilon_i\cdot {\cal J}_{i+1}\cdot k_i\over s_{i(i+1)}}\Big)\,\partial_{\W\epsilon_{i+1}\cdot k_{i-1}}\,\W{\cal P}_1\,{\cal A}_{\rm GR}({\pmb h}_n\setminus i)\,,~~\label{equ-T1-GR-subsub}
\eea
and the solution is found to be
\bea
S=\sum_{j\neq i}\,{(\epsilon_i\cdot {\cal J}_j\cdot k_i)\,(\W\epsilon_i\cdot \W{\cal J}_j\cdot k_i)\over s_{ij}}\,,~~\label{GRs2-solu}
\eea
by using the analogous technic. For Einstein gravity with ${\rm Pf}'\Psi(\epsilon_p,k_p)={\rm Pf}'\W\Psi(\W\epsilon_p,k_p)$, the above formula is reduced to
\bea
S^{(2)_i}_h={1\over2}\,\sum_{j\neq i}\,{(\epsilon_i\cdot J_j\cdot k_i)^2\over s_{ij}}
=-{1\over 2}\,\sum_{j\neq i}\,{k_i\cdot J_j\cdot\varepsilon_i\cdot J_j\cdot k_i\over s_{ij}}\,,~~\label{GRs2-solu-Ein}
\eea
coincide with the result in \cite{Cachazo:2014fwa,Zlotnikov:2014sva}, where the factor $1/2$ is introduced to cancel the over-counting. In the next subsection, we will explain that the solution in \eref{GRs2-solu}
only corresponds to a part of the sub-sub-leading soft behavior of extended gravity which can be detected by the operator ${\cal T}[1,\cdots,n]$, rather than the full one. On the other hand, the formula in \eref{GRs2-solu-Ein} serves as the complete sub-sub-leading soft factor for Einstein gravity.

\subsection{Higher order}
\label{subsec-higherGR}

This subsection aims to argue that the factorized formula \eref{GR-fac} of GR amplitudes, with the expected soft factor in \eref{softfac-condi-GR}, can not be found at the $3^{\rm th}$ order. The argument is similar as that for excluding the expected YM soft factor $S^{(2)_i}_g$, in section \ref{subsec-higherYM}. Paralleled to ${\cal T}_2[1,\cdots,n]$ in \eref{T2-np}, we now choose the operator
\bea
\W{\cal T}_2[1,\cdots,n]=\W{\cal T}_{21}[1,\cdots,n]+\W{\cal T}_{22}[1,\cdots,n]\,,
\eea
where
\bea
\W{\cal T}_{21}[1,\cdots,n]=
\Big(\prod_{a=i+1}^{n-1}\,\partial_{\W\epsilon_a\cdot k_{a-1}}\Big)\,\partial_{\W\epsilon_i\cdot k_{i-2}}\,\Big(\prod_{a=2}^{i-1}\,\partial_{\W\epsilon_a\cdot k_{a-1}}\Big)\,\partial_{\W\epsilon_1\cdot\W\epsilon_n}\,,~~\label{T21-np-GR}
\eea
and
\bea
\W{\cal T}_{22}[1,\cdots,n]=-\partial_{\W\epsilon_{i-1}\cdot k_i}\,
\Big(\prod_{a=i+1}^{n-1}\,\partial_{\W\epsilon_a\cdot k_{a-1}}\Big)\,\partial_{\W\epsilon_i\cdot k_{i-2}}\,\Big(\prod_{a=2}^{i-2}\,\partial_{\W\epsilon_a\cdot k_{a-1}}\Big)\,\partial_{\W\epsilon_1\cdot\W\epsilon_n}\,.~~\label{T22-np-GR}
\eea
It is straightforward to verify that the operator $\W{\cal T}_{21}[1,\cdots,n]$ annihilates ${\cal A}^{(0)_i}_{\rm GR}(\pmb{h}_n)$, while
$\W{\cal T}_{22}[1,\cdots,n]$ annihilates both ${\cal A}^{(0)_i}_{\rm GR}(\pmb{h}_n)$ and ${\cal A}^{(1)_i}_{\rm GR}(\pmb{h}_n)$. The operator $\W{\cal T}_2[1,\cdots,n]$ links the soft behaviors of GR and YM amplitudes as follows
\bea
{\cal A}^{(a)_i}_{\rm YM}(1,\cdots,n)=\W{\cal T}_{21}[1,\cdots,n]\,{\cal A}^{(a+1)_i}_{\rm GR}(\pmb{h}_n)+\W{\cal T}_{22}[1,\cdots,n]\,{\cal A}^{(a+2)_i}_{\rm GR}(\pmb{h}_n)\,.~~\label{transmu-T2-G}
\eea

Before studying the soft behavior of GR amplitudes at the $3^{\rm th}$ order, let us verify that the transmutation relation \eref{transmu-T2-G} holds for $a=0$, namely,
\bea
{\cal A}^{(0)_i}_{\rm YM}(1,\cdots,n)=\W{\cal T}_{21}[1,\cdots,n]\,{\cal A}^{(1)_i}_{\rm GR}(\pmb{h}_n)+\W{\cal T}_{22}[1,\cdots,n]\,{\cal A}^{(2)_i}_{\rm GR}(\pmb{h}_n)\,.~~\label{to-verify}
\eea
The above statement only holds for Einstein gravity, with $S^{(1)_i}_h$ and $S^{(2)_i}_h$ given in \eref{GRs1-solu-Ein} and \eref{GRs2-solu-Ein}.
For the general extended gravity, the transmutation in \eref{to-verify} does not hold if we naively regard the solution in \eref{GRs2-solu} as $S^{(2)_i}_h$. This observation means the sub-sub-leading soft behavior with soft factor in \eref{GRs2-solu} is not the complete one. Therefore, let us restrict our selves to the Einstein gravity.

Since the transmutation operator makes sense for amplitudes of extended gravity, we need to keep notations $\epsilon_p$ and $\W\epsilon_p$ to manifest the double copy structure, and assume that each differential in the combinatorial operator $\W{\cal T}_2[1,\cdots,n]$ only acts on Lorentz invariants which carry $\W\epsilon_p$. For the $\W{\cal T}_{21}[1,\cdots,n]$ part, we realize this goal by going back to the formula of sub-leading soft factor in \eref{GRs1-solu}, which is equivalent to \eref{GRs1-solu-Ein} when setting $\epsilon_p^\mu=\W\epsilon_p^\mu$.
Using the sub-leading soft factor in \eref{GRs1-solu}, we get
\bea
\W{\cal T}_{21}[1,\cdots,n]\,{\cal A}^{(1)_i}_{\rm GR}(\pmb{h}_n)
&=&\partial_{\W\epsilon_{i+1}\cdot k_i}\,\partial_{\W\epsilon_i\cdot k_{i-2}}\,S^{(1)_i}_h\,\W{\cal P}_{21}\,{\cal A}_{\rm GR}(\pmb{h}_n\setminus i)\nn
&=&\Big({2\,\epsilon_i\cdot k_{i-2}\over s_{i(i-2)}}-{\epsilon_i\cdot k_{i+1}\over s_{i(i+1)}}\Big)\,\partial_{\W\epsilon_{i+1}\cdot k_{i-2}}\,\W{\cal P}_{21}\,{\cal A}_{\rm GR}(\pmb{h}_n\setminus i)\,,~~\label{T21-GR-resul}
\eea
where the commutativity in \eref{commu-gener-GR} is used again. In the above, the operator $\W{\cal P}_{21}$ is given as
\bea
\W{\cal P}_{21}=\Big(\prod_{a=i+2}^{n-1}\,\partial_{\W\epsilon_a\cdot k_{a-1}}\Big)\,\Big(\prod_{a=2}^{i-1}\,\partial_{\W\epsilon_a\cdot k_{a-1}}\Big)\,\partial_{\W\epsilon_1\cdot\W\epsilon_n}\,.
\eea
The second equality of \eref{T21-GR-resul} is obtained as follows. The operator $S_2$ in \eref{GRs1-solu} acts on the Lorentz invariant $\W\epsilon_{i+1}\cdot k_{i-2}$ as
\bea
S_2\,(\W\epsilon_{i+1}\cdot k_{i-2})&=&\Big({\epsilon_i\cdot k_{i-2}\over s_{i(i-2)}}-{\epsilon_i\cdot k_{i+1}\over s_{i(i+1)}}\Big)\,\W\epsilon_{i+1}\cdot \W f_i\cdot k_{i-2}\,,
\eea
which includes $(\W\epsilon_{i+1}\cdot k_i)(\W\epsilon_i\cdot k_{i-2})$ in the factor $\W\epsilon_{i+1}\cdot \W f_i\cdot k_{i-2}$, while
the operator $S_1$ in \eref{GRs1-solu} acts on the Lorentz invariant $\W\epsilon_{i+1}\cdot k_{i-2}$ as
\bea
S_1\,(\W\epsilon_{i+1}\cdot k_{i-2})&=&\Big({\W\epsilon_i\cdot k_{i-2}\over s_{i(i-2)}}-{\W\epsilon_i\cdot k_{i+1}\over s_{i(i+1)}}\Big)\,\W\epsilon_{i+1}\cdot f_i\cdot k_{i-2}\,,
\eea
Then the second equality is indicated by the observation that the effect of applying $\partial_{\W\epsilon_{i+1}\cdot k_i}\partial_{\W\epsilon_i\cdot k_{i-2}}$ is turning $(\W\epsilon_{i+1}\cdot k_i)(\W\epsilon_i\cdot k_{i-2})$ to $1$ while eliminating all terms without $(\W\epsilon_{i+1}\cdot k_i)(\W\epsilon_i\cdot k_{i-2})$, and the effect of performing $\partial_{\W\epsilon_{i+1}\cdot k_{i-2}}$ is turning $\W\epsilon_{i+1}\cdot k_{i-2}$ to $1$ while annihilating all terms without $\W\epsilon_{i+1}\cdot k_{i-2}$.

Meanwhile, using the sub-sub-leading soft factor in \eref{GRs2-solu-Ein}, we find
\bea
& &\W{\cal T}_{22}[1,\cdots,n]\,{\cal A}^{(2)_i}_{\rm GR}(\pmb{h}_n)\nn
&=&-\partial_{\W\epsilon_{i+1}\cdot k_i}\,\partial_{\W\epsilon_i\cdot k_{i-2}}\,\partial_{\W\epsilon_{i-1}\cdot k_i}\,S^{(2)_i}_h\,\W{\cal P}_{22}
\,{\cal A}_{\rm GR}(\pmb{h}_n\setminus i)\nn
&=&\Big({\epsilon_i\cdot k_{i+1}\over s_{i(i+1)}}\,\partial_{\W\epsilon_{i+1}\cdot k_{i-2}}\,\partial_{\W\epsilon_{i-1}\cdot k_{i+1}}
+{\epsilon_i\cdot k_{i-1}\over s_{i(i-1)}}\,\partial_{\W\epsilon_{i-1}\cdot k_{i-2}}\,\partial_{\W\epsilon_{i+1}\cdot k_{i-1}}
-{2\,\epsilon_i\cdot k_{i-2}\over s_{i(i-2)}}\,\partial_{\W\epsilon_{i-1}\cdot k_{i-2}}\,\partial_{\W\epsilon_{i+1}\cdot k_{i-2}}\Big)\nn
& &\W{\cal P}_{22}
\,{\cal A}_{\rm GR}(\pmb{h}_n\setminus i)\,,~~\label{T22-GR-resul}
\eea
where
\bea
\W{\cal P}_{22}=\Big(\prod_{a=i+2}^{n-1}\,\partial_{\W\epsilon_a\cdot k_{a-1}}\Big)\,\Big(\prod_{a=2}^{i-2}\,\partial_{\W\epsilon_a\cdot k_{a-1}}\Big)\,\partial_{\W\epsilon_1\cdot\W\epsilon_n}\,.
\eea
The second equality in \eref{T22-GR-resul} is based on the observations
\bea
S^{(2)_i}_h\,(\epsilon_{i+1}\cdot k_{i-2})\,(\epsilon_{i-1}\cdot k_{i+1})&=&-{(\epsilon_{i+1}\cdot f_i\cdot k_{i-2})\,(\epsilon_{i-1}\cdot f_i\cdot k_{i+1})\over s_{i(i+1)}}\nn
&\sim&-{(\W\epsilon_{i+1}\cdot k_i)\,(\W\epsilon_{i-1}\cdot \W k_i)\,(\W\epsilon_i\cdot k_{i-2})\,(\epsilon_i\cdot k_{i+1})\over s_{i(i+1)}}\,,\nn
S^{(2)_i}_h\,(\epsilon_{i-1}\cdot k_{i-2})\,(\epsilon_{i+1}\cdot k_{i-1})&=&-{(\epsilon_{i-1}\cdot  f_i\cdot k_{i-2})\,(\epsilon_{i+1}\cdot f_i\cdot k_{i-1})\over s_{i(i-1)}}\nn
&\sim&-{(\W\epsilon_{i+1}\cdot k_i)\,(\W\epsilon_{i-1}\cdot \W k_i)\,(\W\epsilon_i\cdot k_{i-2})\,(\epsilon_i\cdot k_{i-1})\over s_{i(i-1)}}\,,\nn
S^{(2)_i}_h\,(\epsilon_{i-1}\cdot k_{i-2})\,(\epsilon_{i+1}\cdot k_{i-2})&=&{(\epsilon_{i-1}\cdot  f_i\cdot k_{i-2})\,(\epsilon_{i+1}\cdot f_i\cdot k_{i-2})\over s_{i(i-2)}}\nn
&\sim&{2\,(\W\epsilon_{i+1}\cdot k_i)\,(\W\epsilon_{i-1}\cdot \W k_i)\,(\W\epsilon_i\cdot k_{i-2})\,(\epsilon_i\cdot k_{i-2})\over s_{i(i-2)}}\,.
\eea
Here $\sim$ means collecting effective terms survive under the action of $\partial_{\W\epsilon_{i+1}\cdot k_i}\,\partial_{\W\epsilon_i\cdot k_{i-2}}\,\partial_{\W\epsilon_{i-1}\cdot k_i}$. We turned $\epsilon_{i+1}$ and $\epsilon_{i-1}$ to $\W\epsilon_{i+1}$ and $\W\epsilon_{i-1}$,
based on the fact that $\W{\cal T}_{22}[1,\cdots,n]$ only acts on ${\rm Pf}'\W\Psi(\W\epsilon_p,k_p)$, and kept the structure $\varepsilon^{\mu\nu}_i=\epsilon_i^\mu\W\epsilon_i^\nu$. The factor $2$ in the last line comes from two alternative choices of turning one of two $\epsilon_i\cdot k_{i-2}$ to $\W\epsilon_i\cdot k_{i-2}$.

Combining \eref{T21-GR-resul} and \eref{T22-GR-resul} together, we arrive at
\bea
& &\W{\cal T}_{21}[1,\cdots,n]\,{\cal A}^{(1)_i}_{\rm GR}(\pmb{h}_n)+\W{\cal T}_{22}[1,\cdots,n]\,{\cal A}^{(2)_i}_{\rm GR}(\pmb{h}_n)\nn
&=&\Big({\epsilon_i\cdot k_{i+1}\over s_{i(i+1)}}\,\partial_{\W\epsilon_{i+1}\cdot k_{i-2}}\,\partial_{\W\epsilon_{i-1}\cdot k_{i+1}}
+{\epsilon_i\cdot k_{i-1}\over s_{i(i-1)}}\,\partial_{\W\epsilon_{i-1}\cdot k_{i-2}}\,\partial_{\W\epsilon_{i+1}\cdot k_{i-1}}
-{\epsilon_i\cdot k_{i+1}\over s_{i(i+1)}}\,\partial_{\W\epsilon_{i-1}\cdot k_{i-2}}\,\partial_{\W\epsilon_{i+1}\cdot k_{i-2}}\Big)\nn
& &\W{\cal P}_{22}
\,{\cal A}_{\rm GR}(\pmb{h}_n\setminus i)\nn
&=&\Big({\epsilon_i\cdot k_{i-1}\over s_{i(i-1)}}
-{\epsilon_i\cdot k_{i+1}\over s_{i(i+1)}}\Big)\,{\cal A}_{\rm YM}(1,\cdots,i-1,i+1,\cdots,n)\,.
\eea
In the last step, we have used the observation that both $\partial_{\W\epsilon_{i-1}\cdot k_{i-2}}\partial_{\W\epsilon_{i+1}\cdot k_{i-1}}$
and $(\partial_{\W\epsilon_{i-1}\cdot k_{i+1}}-\partial_{\W\epsilon_{i-1}\cdot k_{i-2}})\partial_{\W\epsilon_{i+1}\cdot k_{i-2}}$ transmutes
the object $\W{\cal P}_{22}{\cal A}_{\rm GR}(\pmb{h}_n\setminus i)$ to the YM amplitude ${\cal A}_{\rm YM}(1,\cdots,i-1,i+1,\cdots,n)$, since the latter one can be interpreted as $\W{\cal I}_{(i-2)(i-1)(i+1)}\W{\cal I}_{(i-2)(i+1)n}$, which inserts the leg $(i+1)$ between $(i-2)$ and $n$, and subsequently insert $(i-1)$ between $(i-2)$ and $(i+1)$. Consequently, the expected transmutation relation \eref{to-verify} is valid.

Then we turn to study the soft factor $S^{(3)_i}_h$ at the $3^{\rm th}$ order. We use the transmutation relation \eref{transmu-T2-G}
with $a=1$, i.e.,
\bea
{\cal A}^{(1)_i}_{\rm YM}(1,\cdots,n)=\W{\cal T}_{21}[1,\cdots,n]\,{\cal A}^{(2)_i}_{\rm GR}(\pmb{h}_n)+\W{\cal T}_{22}[1,\cdots,n]\,{\cal A}^{(3)_i}_{\rm GR}(\pmb{h}_n)\,,~~\label{subsubsub}
\eea
to solve $S^{(3)_i}_h$ by substituting already known $S^{(1)_i}_g$ and $S^{(2)_i}_h$. Since the complete sub-sub-leading soft factor $S^{(2)_i}_h$ in \eref{GRs2-solu-Ein} only makes sense for Einstein gravity, we restrict ourselves to Einstein gravity when considering $S^{(3)_i}_h$. Meanwhile, we still keep the notation $\epsilon_p$ and $\W\epsilon_p$, to manifest the effect of applying operators $\W{\cal T}_{21}[1,\cdots,n]$ and $\W{\cal T}_{22}[1,\cdots,n]$.

Now let us figure out the expression for each block in \eref{subsubsub}.
We can represent ${\cal A}^{(1)_i}_{\rm YM}(1,\cdots,n)$ as
\bea
{\cal A}^{(1)_i}_{\rm YM}(1,\cdots,n)&=&\Big({\epsilon_i\cdot J_{i-1}\cdot k_i\over s_{i(i-1)}}-{\epsilon_i\cdot J_{i+1}\cdot k_i\over s_{i(i+1)}}\Big)\,{\cal A}_{\rm YM}(1,\cdots,i-1,i+1,\cdots,n)\nn
&=&\Big({\epsilon_i\cdot {\cal J}_{i-1}\cdot k_i\over s_{i(i-1)}}-{\epsilon_i\cdot {\cal J}_{i+1}\cdot k_i\over s_{i(i+1)}}\Big)\,\partial_{\W\epsilon_{i+1}\cdot k_{i-1}}\,\partial_{\W\epsilon_{i-1}\cdot k_{i-2}}\,\W{\cal P}_{22}\,{\cal A}_{\rm GR}(\pmb{h}_n\setminus i)\nn
&=&\Big({\epsilon_i\cdot {\cal J}_{i-1}\cdot k_i\over s_{i(i-1)}}-{\epsilon_i\cdot {\cal J}_{i+1}\cdot k_i\over s_{i(i+1)}}\Big)\,\Big(\partial_{\W\epsilon_{i-1}\cdot k_{i-2}}-\partial_{\W\epsilon_{i-1}\cdot k_{i+1}}\Big)\,\partial_{\W\epsilon_{i+1}\cdot k_{i-2}}\,\W{\cal P}_{22}\,{\cal A}_{\rm GR}(\pmb{h}_n\setminus i)\,,~~\label{block1-G}
\eea
where the second and third lines choose two different schemes of insertions, which are $\W I_{(i-1)(i+1)n}\W I_{(i-2)(i-1)n}$ and $\W I_{(i-2)(i-1)(i+1)}\W I_{(i-2)(i+1)n}$ respectively. Both two choices will be considered. On the other hand, by using the sub-sub-leading soft factor in \eref{GRs2-solu-Ein}, we find
\bea
\W{\cal T}_{21}[1,\cdots,n]\,{\cal A}^{(2)_i}_{\rm GR}(\pmb{h}_n)&=&\partial_{\W\epsilon_{i+1}\cdot k_i}\,\partial_{\W\epsilon_i\cdot k_{i-2}}\,S^{(2)_i}_h\,\W{\cal P}_{21}\,{\cal A}_{\rm GR}(\pmb{h}_n\setminus i)\nn
&=&\Big({\epsilon_i\cdot {\cal J}_{i-2}\cdot k_i\over s_{i(i-2)}}-{\epsilon_i\cdot {\cal J}_{i+1}\cdot k_i\over s_{i(i+1)}}\Big)\,\partial_{\W\epsilon_{i+1}\cdot k_{i-2}}\,\partial_{\W\epsilon_{i-1}\cdot k_{i-2}}\,\W{\cal P}_{22}\,{\cal A}_{\rm GR}(\pmb{h}_n\setminus i)\,,~~\label{block2-G}
\eea
since the effective terms survive under the action of $\partial_{\W\epsilon_{i+1}\cdot k_i}\partial_{\W\epsilon_i\cdot k_{i-2}}$
is $\W\epsilon_{i+1}\cdot \W f_i\cdot k_{i+2}$, arises from acting operators $\W\epsilon_i\cdot \W{\cal J}_{i-2}\cdot k_i$ or $\W\epsilon_i\cdot \W{\cal J}_{i+1}\cdot k_i$ on $\W\epsilon_{i+1}\cdot k_{i+2}$. This observation selects $(\epsilon_i\cdot {\cal J}_{i-2}\cdot k_i)(\W\epsilon_i\cdot \W{\cal J}_{i-2}\cdot k_i)/ s_{i(i-2)}$ and $(\epsilon_i\cdot {\cal J}_{i+1}\cdot k_i)(\W\epsilon_i\cdot \W{\cal J}_{i+1}\cdot k_i)/ s_{i(i+1)}$
in \eref{GRs2-solu-Ein}, ultimately yields \eref{block2-G}.

Substituting the second line of \eref{block1-G} and the relation \eref{block2-G} into \eref{subsubsub} leads to the equations
\bea
& &\partial_{\W\epsilon_{i+1}\cdot k_i}\,\partial_{\W\epsilon_i\cdot k_{i-2}}\,\partial_{\W\epsilon_{i-1}\cdot k_i}\,S^{(3)_i}_h\,\W{\cal P}_{22}
\,{\cal A}_{\rm GR}(\pmb{h}_n\setminus i)\nn
&=&\Big[\Big({\epsilon_i\cdot {\cal J}_{i-2}\cdot k_i\over s_{i(i-2)}}-{\epsilon_i\cdot {\cal J}_{i+1}\cdot k_i\over s_{i(i+1)}}\Big)\,\partial_{\W\epsilon_{i+1}\cdot k_{i-2}}\,\partial_{\W\epsilon_{i-1}\cdot k_{i-2}}-\Big({\epsilon_i\cdot {\cal J}_{i-1}\cdot k_i\over s_{i(i-1)}}-{\epsilon_i\cdot {\cal J}_{i+1}\cdot k_i\over s_{i(i+1)}}\Big)\,\partial_{\W\epsilon_{i+1}\cdot k_{i-1}}\,\partial_{\W\epsilon_{i-1}\cdot k_{i-2}}\Big]\nn
& &\W{\cal P}_{22}\,{\cal A}_{\rm GR}(\pmb{h}_n\setminus i)\,.
\eea
Then, similar to the argument at the end of section \ref{subsec-higherYM}, the solution $S^{(3)_i}_h$ which satisfies the expectation in \eref{softfac-condi-GR} is forbidden by the effect of turning the bilinearity on $k_2$ to linearity, or turning the linearity on $k_{i-1}$ to independence. On the other hand, substituting the third line of \eref{block1-G} yields
\bea
& &\partial_{\W\epsilon_{i+1}\cdot k_i}\,\partial_{\W\epsilon_i\cdot k_{i-2}}\,\partial_{\W\epsilon_{i-1}\cdot k_i}\,S^{(3)_i}_h\,\W{\cal P}_{22}
\,{\cal A}_{\rm GR}(\pmb{h}_n\setminus i)\nn
&=&\Big[\Big({\epsilon_i\cdot {\cal J}_{i-2}\cdot k_i\over s_{i(i-2)}}-{\epsilon_i\cdot {\cal J}_{i+1}\cdot k_i\over s_{i(i+1)}}\Big)\,\partial_{\W\epsilon_{i+1}\cdot k_{i-2}}\,\partial_{\W\epsilon_{i-1}\cdot k_{i-2}}-\Big({\epsilon_i\cdot {\cal J}_{i-1}\cdot k_i\over s_{i(i-1)}}-{\epsilon_i\cdot {\cal J}_{i+1}\cdot k_i\over s_{i(i+1)}}\Big)\,\partial_{\W\epsilon_{i-1}\cdot k_{i-2}}\,\partial_{\W\epsilon_{i+1}\cdot k_{i-2}}\nn
& &+\Big({\epsilon_i\cdot {\cal J}_{i-1}\cdot k_i\over s_{i(i-1)}}-{\epsilon_i\cdot {\cal J}_{i+1}\cdot k_i\over s_{i(i+1)}}\Big)\,\partial_{\W\epsilon_{i-1}\cdot k_{i+1}}\,\partial_{\W\epsilon_{i+1}\cdot k_{i-2}}\Big]\,\W{\cal P}_{22}\,{\cal A}_{\rm GR}(\pmb{h}_n\setminus i)\,,
\eea
then the existence of desired $S^{(3)_i}_h$ is forbidden by the effect of turning the bilinearity on $k_2$ to linearity, or turning the linearity on $k_{i+1}$ to independence.
The above argument excludes the existence of expected universal soft factor $S^{(3)_i}_h$.

\section{Summary}
\label{sec-summary}

In this note, with the help of transmutation operators, we reconstruct known soft factors of YM and GR amplitudes, and prove the nonexistence of
higher order soft factor under the constraint of universality. We also clarify that the consistent soft factors $S^{(1)_i}_h$ and $S^{(2)_i}_h$ of GR amplitudes in literatures should be defined for
pure Einstein gravity, rather than for the extended one. This phenomenon is quite natural, since the asymptotic BMS symmetry which predicts the soft behavior of GR amplitudes at leading, and sub-leading orders only makes sense for pure Einstein gravity \cite{Strominger:2013jfa,He:2014laa}. Our method is purely bottom-up thus can not reveal the underlying symmetry, but also leads to the conclusion coincide with the prediction of symmetry.

It is also interesting to relax the requirement of universality, and study soft behavior at higher orders. For example, one can find universal soft factor of BAS amplitudes with the number of external legs $n\geq5$, while the soft behavior of $4$-point amplitudes is distinct. It means one can still talk about the universal soft behavior at a "weaker" level. It is natural to expect the existence of similar feature in YM and GR cases.
An interesting future direction is to figure out such "weaker" universal soft factors, and investigate their physical applications.

\section*{Acknowledgments}

This work is supported by NSFC under Grant No. 11805163.


\bibliographystyle{JHEP}

\bibliography{reference}

\providecommand{\href}[2]{#2}\begingroup\raggedright\begin{thebibliography}{10}

\bibitem{Low:1958sn}
F.~E. Low, {\it {Bremsstrahlung of very low-energy quanta in elementary
  particle collisions}},  {\em Phys. Rev.} {\bf 110} (1958) 974--977.

\bibitem{Weinberg:1965nx}
S.~Weinberg, {\it {Infrared photons and gravitons}},  {\em Phys. Rev.} {\bf
  140} (1965) B516--B524.

\bibitem{Cachazo:2014fwa}
F.~Cachazo and A.~Strominger, {\it {Evidence for a New Soft Graviton Theorem}},
   \href{http://arxiv.org/abs/1404.4091}{{\tt arXiv:1404.4091}}.

\bibitem{Casali:2014xpa}
E.~Casali, {\it {Soft sub-leading divergences in Yang-Mills amplitudes}},  {\em
  JHEP} {\bf 08} (2014) 077, [\href{http://arxiv.org/abs/1404.5551}{{\tt
  arXiv:1404.5551}}].

\bibitem{Schwab:2014xua}
B.~U.~W. Schwab and A.~Volovich, {\it {Subleading Soft Theorem in Arbitrary
  Dimensions from Scattering Equations}},  {\em Phys. Rev. Lett.} {\bf 113}
  (2014), no.~10 101601, [\href{http://arxiv.org/abs/1404.7749}{{\tt
  arXiv:1404.7749}}].

\bibitem{Afkhami-Jeddi:2014fia}
N.~Afkhami-Jeddi, {\it {Soft Graviton Theorem in Arbitrary Dimensions}},
  \href{http://arxiv.org/abs/1405.3533}{{\tt arXiv:1405.3533}}.

\bibitem{Zlotnikov:2014sva}
M.~Zlotnikov, {\it {Sub-sub-leading soft-graviton theorem in arbitrary
  dimension}},  {\em JHEP} {\bf 10} (2014) 148,
  [\href{http://arxiv.org/abs/1407.5936}{{\tt arXiv:1407.5936}}].

\bibitem{Britto:2004ap}
R.~Britto, F.~Cachazo, and B.~Feng, {\it {New recursion relations for tree
  amplitudes of gluons}},  {\em Nucl. Phys. B} {\bf 715} (2005) 499--522,
  [\href{http://arxiv.org/abs/hep-th/0412308}{{\tt hep-th/0412308}}].

\bibitem{Britto:2005fq}
R.~Britto, F.~Cachazo, B.~Feng, and E.~Witten, {\it {Direct proof of tree-level
  recursion relation in Yang-Mills theory}},  {\em Phys. Rev. Lett.} {\bf 94}
  (2005) 181602, [\href{http://arxiv.org/abs/hep-th/0501052}{{\tt
  hep-th/0501052}}].

\bibitem{Cachazo:2013gna}
F.~Cachazo, S.~He, and E.~Y. Yuan, {\it {Scattering equations and
  Kawai-Lewellen-Tye orthogonality}},  {\em Phys. Rev.} {\bf D90} (2014), no.~6
  065001, [\href{http://arxiv.org/abs/1306.6575}{{\tt arXiv:1306.6575}}].

\bibitem{Cachazo:2013hca}
F.~Cachazo, S.~He, and E.~Y. Yuan, {\it {Scattering of Massless Particles in
  Arbitrary Dimensions}},  {\em Phys. Rev. Lett.} {\bf 113} (2014), no.~17
  171601, [\href{http://arxiv.org/abs/1307.2199}{{\tt arXiv:1307.2199}}].

\bibitem{Cachazo:2013iea}
F.~Cachazo, S.~He, and E.~Y. Yuan, {\it {Scattering of Massless Particles:
  Scalars, Gluons and Gravitons}},  {\em JHEP} {\bf 07} (2014) 033,
  [\href{http://arxiv.org/abs/1309.0885}{{\tt arXiv:1309.0885}}].

\bibitem{Cachazo:2014nsa}
F.~Cachazo, S.~He, and E.~Y. Yuan, {\it {Einstein-Yang-Mills Scattering
  Amplitudes From Scattering Equations}},  {\em JHEP} {\bf 01} (2015) 121,
  [\href{http://arxiv.org/abs/1409.8256}{{\tt arXiv:1409.8256}}].

\bibitem{Cachazo:2014xea}
F.~Cachazo, S.~He, and E.~Y. Yuan, {\it {Scattering Equations and Matrices:
  From Einstein To Yang-Mills, DBI and NLSM}},  {\em JHEP} {\bf 07} (2015) 149,
  [\href{http://arxiv.org/abs/1412.3479}{{\tt arXiv:1412.3479}}].

\bibitem{Bern:2014oka}
Z.~Bern, S.~Davies, and J.~Nohle, {\it {On Loop Corrections to Subleading Soft
  Behavior of Gluons and Gravitons}},  {\em Phys. Rev. D} {\bf 90} (2014),
  no.~8 085015, [\href{http://arxiv.org/abs/1405.1015}{{\tt arXiv:1405.1015}}].

\bibitem{He:2014bga}
S.~He, Y.-t. Huang, and C.~Wen, {\it {Loop Corrections to Soft Theorems in
  Gauge Theories and Gravity}},  {\em JHEP} {\bf 12} (2014) 115,
  [\href{http://arxiv.org/abs/1405.1410}{{\tt arXiv:1405.1410}}].

\bibitem{Cachazo:2014dia}
F.~Cachazo and E.~Y. Yuan, {\it {Are Soft Theorems Renormalized?}},
  \href{http://arxiv.org/abs/1405.3413}{{\tt arXiv:1405.3413}}.

\bibitem{Bianchi:2014gla}
M.~Bianchi, S.~He, Y.-t. Huang, and C.~Wen, {\it {More on Soft Theorems: Trees,
  Loops and Strings}},  {\em Phys. Rev. D} {\bf 92} (2015), no.~6 065022,
  [\href{http://arxiv.org/abs/1406.5155}{{\tt arXiv:1406.5155}}].

\bibitem{Sen:2017nim}
A.~Sen, {\it {Subleading Soft Graviton Theorem for Loop Amplitudes}},  {\em
  JHEP} {\bf 11} (2017) 123, [\href{http://arxiv.org/abs/1703.00024}{{\tt
  arXiv:1703.00024}}].

\bibitem{Nguyen:2009jk}
D.~Nguyen, M.~Spradlin, A.~Volovich, and C.~Wen, {\it {The Tree Formula for MHV
  Graviton Amplitudes}},  {\em JHEP} {\bf 07} (2010) 045,
  [\href{http://arxiv.org/abs/0907.2276}{{\tt arXiv:0907.2276}}].

\bibitem{Boucher-Veronneau:2011rwd}
C.~Boucher-Veronneau and A.~J. Larkoski, {\it {Constructing Amplitudes from
  Their Soft Limits}},  {\em JHEP} {\bf 09} (2011) 130,
  [\href{http://arxiv.org/abs/1108.5385}{{\tt arXiv:1108.5385}}].

\bibitem{Rodina:2018pcb}
L.~Rodina, {\it {Scattering Amplitudes from Soft Theorems and Infrared
  Behavior}},  {\em Phys. Rev. Lett.} {\bf 122} (2019), no.~7 071601,
  [\href{http://arxiv.org/abs/1807.09738}{{\tt arXiv:1807.09738}}].

\bibitem{Ma:2022qja}
S.~Ma, R.~Dong, and Y.-J. Du, {\it {Constructing EYM amplitudes by inverse soft
  limit}},  {\em JHEP} {\bf 05} (2023) 196,
  [\href{http://arxiv.org/abs/2211.10047}{{\tt arXiv:2211.10047}}].

\bibitem{Cheung:2014dqa}
C.~Cheung, K.~Kampf, J.~Novotny, and J.~Trnka, {\it {Effective Field Theories
  from Soft Limits of Scattering Amplitudes}},  {\em Phys. Rev. Lett.} {\bf
  114} (2015), no.~22 221602, [\href{http://arxiv.org/abs/1412.4095}{{\tt
  arXiv:1412.4095}}].

\bibitem{Cheung:2015ota}
C.~Cheung, K.~Kampf, J.~Novotny, C.-H. Shen, and J.~Trnka, {\it {On-Shell
  Recursion Relations for Effective Field Theories}},  {\em Phys. Rev. Lett.}
  {\bf 116} (2016), no.~4 041601.

\bibitem{Luo:2015tat}
H.~Luo and C.~Wen, {\it {Recursion relations from soft theorems}},  {\em JHEP}
  {\bf 03} (2016) 088, [\href{http://arxiv.org/abs/1512.06801}{{\tt
  arXiv:1512.06801}}].

\bibitem{Elvang:2018dco}
H.~Elvang, M.~Hadjiantonis, C.~R.~T. Jones, and S.~Paranjape, {\it {Soft
  Bootstrap and Supersymmetry}},  {\em JHEP} {\bf 01} (2019) 195,
  [\href{http://arxiv.org/abs/1806.06079}{{\tt arXiv:1806.06079}}].

\bibitem{Zhou:2022orv}
K.~Zhou, {\it {Tree level amplitudes from soft theorems}},  {\em JHEP} {\bf 03}
  (2023) 021, [\href{http://arxiv.org/abs/2212.12892}{{\tt arXiv:2212.12892}}].

\bibitem{Wei:2023yfy}
F.-S. Wei and K.~Zhou, {\it {Expanding single-trace YMS amplitudes with
  gauge-invariant coefficients}},  {\em Eur. Phys. J. C} {\bf 84} (2024), no.~1
  29, [\href{http://arxiv.org/abs/2306.14774}{{\tt arXiv:2306.14774}}].

\bibitem{Hu:2023lso}
C.~Hu and K.~Zhou, {\it {Recursive construction for expansions of tree
  Yang\textendash{}Mills amplitudes from soft theorem}},  {\em Eur. Phys. J. C}
  {\bf 84} (2024), no.~3 221, [\href{http://arxiv.org/abs/2311.03112}{{\tt
  arXiv:2311.03112}}].

\bibitem{Du:2024dwm}
Y.-J. Du and K.~Zhou, {\it {Multi-trace YMS amplitudes from soft behavior}},
  {\em JHEP} {\bf 03} (2024) 081, [\href{http://arxiv.org/abs/2401.03879}{{\tt
  arXiv:2401.03879}}].

\bibitem{Cheung:2017ems}
C.~Cheung, C.-H. Shen, and C.~Wen, {\it {Unifying Relations for Scattering
  Amplitudes}},  {\em JHEP} {\bf 02} (2018) 095,
  [\href{http://arxiv.org/abs/1705.03025}{{\tt arXiv:1705.03025}}].

\bibitem{Zhou:2018wvn}
K.~Zhou and B.~Feng, {\it {Note on differential operators, CHY integrands, and
  unifying relations for amplitudes}},  {\em JHEP} {\bf 09} (2018) 160,
  [\href{http://arxiv.org/abs/1808.06835}{{\tt arXiv:1808.06835}}].

\bibitem{Bollmann:2018edb}
M.~Bollmann and L.~Ferro, {\it {Transmuting CHY formulae}},  {\em JHEP} {\bf
  01} (2019) 180, [\href{http://arxiv.org/abs/1808.07451}{{\tt
  arXiv:1808.07451}}].

\bibitem{Strominger:2013jfa}
A.~Strominger, {\it {On BMS Invariance of Gravitational Scattering}},  {\em
  JHEP} {\bf 07} (2014) 152, [\href{http://arxiv.org/abs/1312.2229}{{\tt
  arXiv:1312.2229}}].

\bibitem{He:2014laa}
T.~He, V.~Lysov, P.~Mitra, and A.~Strominger, {\it {BMS supertranslations and
  Weinberg\textquoteright{}s soft graviton theorem}},  {\em JHEP} {\bf 05}
  (2015) 151, [\href{http://arxiv.org/abs/1401.7026}{{\tt arXiv:1401.7026}}].

\end{thebibliography}\endgroup

\end{document}